\documentclass[aps, prc, nofootinbib, superscriptaddress, preprint]{revtex4-2}

\usepackage[utf8]{inputenc}
\usepackage{graphicx}
\usepackage[caption=false]{subfig}
\usepackage[dvipsnames]{xcolor}
\usepackage{amsthm, amsfonts, amsmath, mathrsfs}
\usepackage{bbm}
\usepackage{slashed}
\usepackage{physics}
\usepackage{xspace}
\usepackage{enumitem}
\setlist[enumerate]{itemsep=0.01cm}
\usepackage{orcidlink}
\usepackage{comment}
\usepackage{glossaries}
\setacronymstyle{long-short}
\glsdisablehyper

\usepackage{hyperref}
\hypersetup{colorlinks=true,
    linkcolor=magenta,
    citecolor=blue,
    urlcolor=cyan,
    pdfpagemode=FullScreen,}

\newacronym{pionless}{EFT$_\slashed \pi$}{pionless effective field theory}
\newacronym{eft}{EFT}{effective field theory}
\newacronym{lecs}{LECs}{low energy coefficients}
\newacronym{qed}{QED}{quantum electrodynamics}
\newacronym{qcd}{QCD}{quantum chromodynamics}
\newacronym{latticeqcd}{LQCD}{lattice QCD}
\newacronym{nrqed}{NRQED}{nonrelativistic QED}
\newacronym{vnrqed}{vNRQED}{velocity NRQED}
\newacronym{nrqcd}{NRQCD}{nonrelativistic QCD}
\newacronym{ChiPT}{\ensuremath{\chi}PT}{chiral perturbation theory}
\newacronym{ChiEFT}{\ensuremath{\chi}EFT}{chiral effective field theory}
\newacronym{BSM}{BSM}{Beyond the Standard Model}
\newacronym{NN}{\ensuremath{N\!N}\xspace}{nucleon-nucleon}
\newacronym{vRG}{vRG}{velocity renormalization group}
\newacronym{lo}{LO\ensuremath{_\slashed \pi}}{leading order}
\newacronym{nlo}{\ensuremath{\text{NLO}_{\slashed \pi}}}{next-to-leading order}
\newacronym{n2lo}{\ensuremath{\text{N}^2\text{LO}_{\slashed \pi}}}{next-to-next-to-leading order}
\newacronym{n3lo}{\ensuremath{\text{N}^3\text{LO}_{\slashed \pi}}}{next-to-next-to-next-to-leading order}
\newacronym{n4lo}{\ensuremath{\text{N}^4\text{LO}}}{next-to-next-to-next-to-next-to-leading order}
\newacronym{dimreg}{DimReg}{dimensional regularization}
\newacronym{av18}{AV18}{Argonne \ensuremath{v18}}
\newacronym{LLalpha}{\ensuremath{\text{LL}_\alpha}}{leading-logarithm-in \ensuremath{\alpha}}
\newacronym{NLLalpha}{\ensuremath{\text{NLL}_\alpha}}{next-to-leading-logarithm-in \ensuremath{\alpha}}
\newacronym{PDS}{PDS}{Power Divergence Subtraction}
\newacronym{BBN}{BBN}{big bang nucleosynthesis}

\newcommand{\oneS}{{{}^{1}\!S_0}}
\newcommand{\threeS}{{{}^{3}\!S_1}}

\newcommand{\del}{\nabla}

\newcommand{\calO}{\ensuremath{\mathcal{O}}}

\newcommand{\calL}{\ensuremath{\mathcal{L}}}

\newcommand{\beq}{\begin{equation}}
\newcommand{\eeq}{\end{equation}}

\newcommand{\NN}{\ensuremath{N\!N}\xspace}

\begin{document}

\title{Renormalization group analysis of electromagnetic properties of the deuteron}

\author{Thomas R.~Richardson\,\orcidlink{0000-0001-6314-7518}}
\email{thomas.richardson@berkeley.edu}
\affiliation{Institut f\"ur Kernphysik and PRISMA$^+$ Cluster of Excellence, Johannes Gutenberg-Universit\"at, 55128 Mainz, Germany}
\affiliation{Department of Physics, University of California, Berkeley, CA 94720, USA}
\affiliation{Nuclear Science Division, Lawrence Berkeley National Laboratory, Berkeley, CA 94720, USA}

\author{Immo C. Reis\,\orcidlink{0009-0008-3566-6095}}
\email{imreis@students.uni-mainz.de}
\affiliation{Institut f\"ur Kernphysik and PRISMA$^+$ Cluster of Excellence, Johannes Gutenberg-Universit\"at, 55128 Mainz, Germany}

\begin{abstract}
    The role of radiative corrections in low energy nuclear physics is beginning to receive more scrutiny.
    We examine the impact of these corrections for the deuteron charge form factor and the radiative capture process $np \to d \gamma$ through the velocity renormalization group.
    In both cases, we find percent level shifts in the relevant observables after evolving the subtraction velocity to the typical velocity of nucleons in the bound state.
    This suggests that electromagnetic corrections constitute a non-negligible source of uncertainty in existing few-body calculations.
\end{abstract}

\maketitle

\newpage


\section{Introduction}
    \label{sec:introduction}

In recent years, precision electroweak structure in few-nucleon systems has received significant experimental, observational, and theoretical attention.
Specifically, we have in mind measurements of charge radii in muonic atom spectroscopy \cite{pohlSizeProton2010, pohlLaserSpectroscopyMuonic2016,antogniniMuonicAtomSpectroscopyImpact2022, ohayonPrecisionMuonicXray2024} and the observation of light element abundances from \gls{BBN} \cite{wagonerSynthesisElementsVery1967, steigmanPrimordialNucleosynthesisPrecision2007, ioccoPrimordialNucleosynthesisPrecision2009, cyburtBigBangNucleosynthesis2016, aghanimPlanck2018Results2020, cookeOnePercentDetermination2018, burlesSharpeningPredictionsBigBang1999}.
These processes have the potential to shed light on new physics, provided that all Standard Model effects are theoretically under control.
In particular, the increasing experimental and observational precision dictates a more thorough understanding of radiative corrections from \gls{qed} in low-energy few-nucleon systems.

Over the course of the last three decades, the most prominent tools to study the electroweak structure of nuclei have been rooted in \gls{eft}.
Many studies are based on \gls{ChiEFT} \cite{weinbergNuclearForcesChiral1990, weinbergEffectiveChiralLagrangians1991, weinbergThreeBodyInteractions1992}, the multi-nucleon extension of \gls{ChiPT}.
There is also the so-called \gls{pionless} \cite{kaplanNucleonNucleonScattering1996, kaplanNewExpansionNucleonnucleon1998, kaplanTwoNucleonSystems1998, vankolckEffectiveFieldTheory1999}, which is suitable for light nuclei where the typical nucleon momentum $|\vb p|$ is much less than the pion mass $m_\pi$ (see Refs.~\cite{bedaqueEffectiveFieldTheory2002, hammerNuclearEffectiveField2020} for reviews).
While both \gls{ChiEFT} and \gls{pionless} in conjunction with modern uncertainty quantification tools \cite{acharyaUncertaintyQuantificationElectromagnetic2023, wesolowskiBayesianParameterEstimation2016, schindlerBayesianMethodsParameter2009, furnstahlQuantifyingTruncationErrors2015, epelbaumImprovedChiralNucleonnucleon2015} have led to more and more sophisticated analyses in low-energy nuclear physics, they are currently incapable of teasing apart the different Standard Model effects since the parameters of the \gls{eft}s have been fit to experimental data rather than fixing them to \gls{qcd} results. 
Moreover, there is a noticeable lack of a dedicated study of radiative corrections apart from a few works \cite{vankolckElectromagneticCorrectionsOnePionExchange1998, ciriglianoInitioElectroweakCorrections2024, ciriglianoRadiativeCorrectionsSuperallowed2024}.
Yet, we know that in order to produce the correct infrared structure of the full theory, i.e. \gls{qcd} plus \gls{qed} and any higher dimensional operators, all relevant degrees of freedom that can go on-shell in the low-energy regime must be retained.
In the low-energy regime where \gls{ChiEFT} and \gls{pionless} are valid, this implies that the correct \gls{eft} should have dynamical photons that are near the mass shell.
In fact, these issues seem to be coming to the forefront of nuclear physics in the context of superallowed $\beta$ decay \cite{ciriglianoInitioElectroweakCorrections2024, ciriglianoRadiativeCorrectionsSuperallowed2024}, for example.

In the same timeframe that led to the development of nuclear \gls{eft}s, similar \gls{eft}s were developed specifically to handle radiative corrections in nonrelativstic \gls{qcd} and \gls{qed} bound states \cite{caswellEffectiveLagrangiansBound1986, labelleEffectiveFieldTheories1998, lukeBoundStatesPower1997, lukePowerCountingDimensionally1998, griesshammerThresholdExpansionDimensionally1998, griesshammerPowerCountingBeta2000, grinsteinEffectiveFieldTheory1998, lukeRenormalizationGroupScaling2000, manoharQCDHeavyquarkPotential2000, manoharLogarithms$ensuremathalpha$QED2000, manoharRenormalizationGroupAnalysis2000, manoharRenormalizationGroupCorrelated2000, manoharRunningHeavyQuark2001, manoharZeroBinModeFactorization2007, pinedaEffectiveFieldTheory1998, pinedaLambShiftDimensional1998, pinedaMatchingOneLoop1998, pinedaPotentialNRQEDPositronium1998, brambillaEffectivefieldTheoriesHeavy2005, hoangRenormalizationGroupImprovedCalculation2001, hoangRunningCoulombPotential2001, hoangUltrasoftRenormalizationNonrelativistic2003}.
Borrowing techniques from \gls{nrqcd}, in particular the \gls{vRG} \cite{lukeRenormalizationGroupScaling2000}, we initiated a study of the role of radiative corrections relevant for light nuclei \cite{richardsonRadiativeCorrectionsRenormalization2024}.
There, we derived the leading anomalous dimension of the \gls{NN} potential and used the \gls{vRG} to sum the \gls{LLalpha} series into the \gls{nlo} and \gls{n2lo} \gls{lecs} in the standard \gls{pionless} power counting.
Subsequently, we saw a few-percent shift to the deuteron binding energy at \gls{n2lo}+\gls{LLalpha} through renormalization group improved perturbation theory.
In this work, we extend the technique to the study of the electromagnetic properties of the deuteron. 
We will focus on the charge form factor at \gls{n2lo}+\gls{LLalpha} relevant for the extraction of the charge radius and the radiative neutron capture process $np \to d \gamma$ at \gls{nlo}+\gls{LLalpha} relevant for precision cosmology.

In Sec.~\ref{sec:background}, we will recapitulate the construction of the effective Lagrangian and the renormalization of the \NN interaction.
We discuss the results for the form factor in Sec.~\ref{sec:charge_ff} and for the radiative capture cross section in Sec.~\ref{sec:npdgamma}.
Finally, we summarize our results in Sec.~\ref{sec:conclusion}.


\section{The effective Lagrangian and ultrasoft renormalization}
    \label{sec:background}

Nonrelativistic \gls{qcd} and \gls{qed} first emerged in the mid-1980's \cite{caswellEffectiveLagrangiansBound1986}.
Initially, the full gauge theory with relativistic fermions was matched onto a theory with gauge bosons and nonrelativistic fermions of mass $m$ with momentum $\vb p$ and energy $E \sim \vb p^2/m$.
If we introduce the velocity $v = | \vb p |/m$, then we can distinguish several relevant  (energy, momentum) scales:
    \begin{enumerate}
        \item hard $\sim (m,m)$ 
        \item soft $\sim (m v, m v)$
        \item ultrasoft $\sim (m v^2, m v^2)$
        \item potential $\sim (m v^2, m v)$ \, .
    \end{enumerate}
In a loop diagram, a virtual photon can couple to different scales simultaneously.
This is undesirable from the \gls{eft} perspective since it prohibits a homogenous power counting, i.e., any given Feynman diagram does not scale with a unique power of $v$.
It was demonstrated in Ref.~\cite{labelleEffectiveFieldTheories1998} that performing a multipole expansion of the ultrasoft modes resolves this power counting issue.
In Ref.~\cite{grinsteinEffectiveFieldTheory1998}, it was first shown how to implement the multipole expansion at the level of the Lagrangian while a velocity power counting rule was developed in Ref.~\cite{lukeBoundStatesPower1997}.
Afterwards, these ideas were implemented in \gls{nrqcd} \cite{lukePowerCountingDimensionally1998}.
However, this work only contained potential and ultrasoft gluons.
The importance of soft gluons was pointed out in Refs.~\cite{griesshammerThresholdExpansionDimensionally1998, griesshammerPowerCountingBeta2000}, which is an implementation of the threshold expansion \cite{benekeAsymptoticExpansionFeynman1998} at the level of the Lagrangian.

At this point, two separate developments occurred.
On one hand there was the development of potential \gls{nrqed} \cite{pinedaEffectiveFieldTheory1998, pinedaLambShiftDimensional1998, pinedaMatchingOneLoop1998, pinedaEffectiveFieldTheory1998, pinedaPotentialNRQEDPositronium1998} (see Ref.~\cite{brambillaEffectivefieldTheoriesHeavy2005} for a review).
In potential \gls{nrqed}, \gls{qed} is matched onto a nonrelativistic Lagrangian at the hard scale.
Then, the renormalization scale is run down to the soft scale where the soft photons are integrated out.
On the other hand, there is \gls{vnrqed} \cite{lukeRenormalizationGroupScaling2000} in which the soft modes are not integrated out at all.
Rather, one begins with a Lagrangian at the hard scale and uses a formulation of the renormalization group referred to as the \gls{vRG} to simultaneously run soft modes from $\mu = m$ to $\mu = mv$ and ultrasoft modes to $\mu = mv^2$, where $\mu$ is the scale introduced in dimensional regularization.
These techniques were subsequently refined and used for a wide variety of applications for both nonrelativistic QED and QCD bound states \cite{manoharLogarithms$ensuremathalpha$QED2000, manoharRenormalizationGroupAnalysis2000, manoharQCDHeavyquarkPotential2000, manoharRenormalizationGroupCorrelated2000, manoharRunningHeavyQuark2001, hoangRenormalizationGroupImprovedCalculation2001, hoangRunningCoulombPotential2001,hoangUltrasoftRenormalizationNonrelativistic2003}.

Now, we can translate these ideas into the framework of \gls{pionless} following Refs.~\cite{lukeBoundStatesPower1997, richardsonRadiativeCorrectionsRenormalization2024}.
The (energy, momentum) scales of \gls{pionless} with virtual photons are:
    \begin{enumerate}
        \item hard $\sim (m_\pi,m_\pi)$ 
        \item soft $\sim (M_N v, M_N v)$
        \item ultrasoft $\sim (M_N v^2, M_N v^2)$
        \item potential $\sim (M_N v^2, M_N v)$
    \end{enumerate}
where $m_\pi$ is the pion mass and $M_N$ is the nucleon mass.
There are different active degrees of freedom that live in each of these regions.
Everything at or above the hard scale is integrated out.
In the potential regime, there are only nucleon fields.
The nonrelativistic four-momentum $P$ of the nucleon is decomposed into a soft piece and a residual ultrasoft piece according to \cite{lukeRenormalizationGroupScaling2000}
    \begin{equation}
            \label{eq:background:nrqed:nucleon_momentum}
        P = (0, \vb p) + (k_0, \vb k) \, ,
    \end{equation}
where $\vb p \sim M_N v$ is the soft component of the nucleon momentum and $k \sim M_N v^2$ is the residual four-momentum of a nucleon on the ultrasoft scale.
With this momentum binning, the nucleon fields can be denoted as $N_{\vb p} (x)$ where $\vb p$ is a soft label, $x$ is the Fourier conjugate of the residual momentum $k$, and $N$ is an isodoublet of the proton and neutron.

The ordinary photon field is split into a mode that lives in the soft region and a mode that lives in the ultrasoft region.
The soft mode $A^\mu_p(x)$ has a soft label of four-momentum $p$ and $x$ is again the Fourier conjugate of the residual momentum $k$.
The ultrasoft mode $A^\mu(x)$ only carries ultrasoft four-momentum $k$, which is again the Fourier conjugate variable to $x$.
Notice that there are no potential photons in the theory.
Since particles in the potential regime have (energy, momentum)$\sim (M_N v^2, M_N v)$, a photon in the potential regime would be far off-shell and can therefore  be integrated out.
Additionally, soft nucleons could be included in the theory, but they can just as well be integrated out for the same reason.
Greater detail concerning this separation of modes and momentum binning can be found in the Appendix.

In order to implement the \gls{vRG} in dimensional regularization, we introduce a soft renormalization scale $\mu_S$ and an ultrasoft renormalization scale $\mu_U$ \cite{lukeRenormalizationGroupScaling2000,manoharRenormalizationGroupCorrelated2000}.
The two scales are correlated according to $\mu_U = \mu_S^2/M_N$ through the introduction of a subtraction velocity $\nu$ according to $\mu_S = M_N \nu$.
The ultrasoft and soft anomalous dimensions of a generic coupling $C$ are obtained through
\cite{lukeRenormalizationGroupScaling2000,manoharRenormalizationGroupCorrelated2000}
    \begin{align}
        \mu_U \frac{d C}{d \mu_U} & = \gamma_U \, , \\
        \mu_S \frac{d C}{d \mu_S} & = \gamma_S \, ,
    \end{align}
which lead to the \gls{vRG} equation
    \begin{align}
        \nu \frac{d C}{d \nu} & = \gamma_S + 2 \gamma_U \, .
    \end{align}
In \gls{nrqed}, the electron mass is integrated out, so the fine-structure constant $\alpha$, which will appear in the anomalous dimensions, does not run.
In the \NN system, the electron is kept as a light degree of freedom, which leads to a more interesting \gls{vRG} structure because $\alpha$ runs.
Here, the anomalous dimensions have a dual expansion,
    \begin{align}
        \gamma & = \sum_{m=-1, n=0} \gamma^{(m, n)} \, ,
    \end{align}
where $\gamma^{(m, n)} \sim v^m \alpha^n$.

\subsection{Single-nucleon sector}
In the one-nucleon sector, ultrasoft photons play a role similar to gluons in heavy quark effective theory \cite{eichtenStaticEffectiveField1990} or pions in heavy baryon \gls{ChiPT} \cite{jenkinsBaryonChiralPerturbation1991}.
Through $O(1/M_N^2)$, the terms in the Lagrangian relevant for this work are
    \begin{align}
        \calL_{N} & = \sum_{\vb p} N^\dagger_{\vb p} \left[ i D_0 + \frac{ \left( \vb p - i \vb D \right)^2}{2 M_N} + \frac{e}{2 M_N} \left(\kappa_0 + \kappa_1 \tau^3 \right) \vec \sigma \cdot \vb B + \frac{e}{8 M_N^2} \left( c_D^{(s)} + c_D^{(v)} \tau^3 \right) \del \cdot \vb E \right] N_{\vb p} \, , \label{eq:single_nucleon_L}
    \end{align}
where $\vec \sigma$ is the vector of Pauli matrices.
The covariant derivative acting on the nucleon field is $D_\mu = \partial_\mu + i e Q_N A_\mu$, where $Q_N = \text{diag} \left( 1, \, 0 \right)$ is the nucleon charge matrix, and the photon field is ultrasoft.
The kinetic term for the nucleon is $\left( \vb p - i \vb D \right)^2$, and its form is fixed by reparameterization invariance \cite{lukeRenormalizationGroupScaling2000}. 
This term should be expanded while keeping only the $\vb p^2$ term in the leading order propagator.
Terms containing factors of $\vb p \cdot \vb D$ or $\vb D^2$ are treated as perturbations; this organization is equivalent to the multipole expansion mentioned above.
The \gls{lecs} $\kappa_0$ ($\kappa_1$) and $c_D^{(s)}$ ($c_D^{(v)}$) are the isoscalar (isovector) anomalous magnetic moment and Darwin terms, respectively.
To obtain a result at \gls{LLalpha}, the \gls{lecs} of this sector could be determined by matching to a \gls{ChiPT} calculation at tree-level with respect to the photon, i.e.~at $O(e)$, at the scale $\nu = m_\pi/M_N$ or possibly from lattice \gls{qcd} calculations.

\begin{figure}
    \centering
    \includegraphics[width=0.3\linewidth]{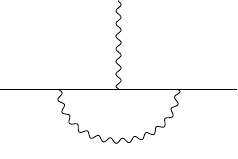}
    \caption{One loop \gls{qed} correction to the nucleon charge form factor}
    \label{fig:proton_ff}
\end{figure}

At the order we are considering, only the proton field strength is renormalized. 
The wave-function renormalization is given by (see Ref.~\cite{manoharHeavyQuarkPhysics2000} and references therein),
    \begin{align}
        Z_p & = 1 + \frac{\alpha(\mu_{US})}{2 \pi \epsilon} \, .
    \end{align}
where the running \gls{qed} coupling in minimal subtraction is 
    \begin{align}
        \alpha(\mu_{US}) & = \frac{\alpha_{\text{OS}}}{1 - \frac{2 \alpha_{\text{OS}}}{3 \pi} \log \left( \frac{\mu_{US}}{m_e} \right)} \, ,
    \end{align}
where $m_e = 511$ keV, $\alpha_{\text{OS}} = 1/137$ with OS indicating that this is the fine structure constant in the on-shell scheme, and $\mu_{US} = M_N \nu^2$ is the ultrasoft subtraction scale.
In Sec.~\ref{sec:charge_ff}, we will reevaluate the deuteron charge form factor through \gls{n2lo}.
At this order, the form factor receives a contribution from the operator typically written as \cite{phillipsImprovingConvergenceEffective2000, lenskyForwardDoublyvirtualCompton2021}
    \begin{align}
        \calL & \supset \frac{1}{6} \del \cdot \vb E N^\dagger\left( r_0^2 + r_1^2 \tau^3 \right)  N \, ,
    \end{align}
where $r_0$ ($r_1$) is referred to as the isoscalar (isovector) nucleon charge radius.
These coefficients are usually fit to the charge radius of the nucleon defined according to
    \begin{equation}
        r_N^2 = 6 \frac{d G_E^{(N)}(q^2)}{d q^2} \big |_{q^2 = 0} \, ,
    \end{equation}
where $G_E^{(N)}$ is the electric Sachs form factor of the nucleon.
This definition of the charge radius is the one most used in measurements of charge radii from spectroscopy or lepton scattering.
However, it is well known that the one-loop \gls{qed} correction to the nucleon form factor shown in Fig.~\ref{fig:proton_ff} is infrared divergent, which makes the interpretation of the coefficient in the Lagrangian as the charge radius complicated \cite{pachuckiRadiativeRecoilCorrection1995, pinedaChiralStructureLamb2005, pesetProtonRadiusPuzzle2021,manoharHeavyQuarkEffective1997}\footnote{At the order we are working, the Darwin term for the neutron does not run. It can therefore reasonably be taken from experiment or lattice \gls{qcd}.}.

In the context of an EFT, the running \gls{lecs} should be considered as matching coefficients that have power series expansions in $\alpha$ \cite{pinedaChiralStructureLamb2005, pesetProtonRadiusPuzzle2021}.
Therefore, we find it more appropriate to work with $c_D$ from Eq.~\eqref{eq:single_nucleon_L}, which can be more clearly matched to the single-nucleon matrix element of the quark electromagnetic current.
For the proton, we have
    \begin{align}
        \bra{\vb p'} \bar q Q_{\text{em}} \gamma^\mu q \ket{\vb p} & = \bar \psi (p') \left[ F^{(p)}_1(q^2) \gamma^\mu + \frac{i}{2 M_N} \sigma^{\mu \nu} q_\nu F^{(p)}_2 (q^2) \right] \psi(p) \label{eq:proton_matrix_element} \, ,
    \end{align}
where $\mu$ is now a Lorentz index, $q = (u, d)^T$ is the isodoublet of up and down quarks, $Q_{\text{em}} = \text{diag} \left(2/3, -1/3 \right)$ is the quark charge matrix, $p$ ($p'$) is the four-momentum of the incoming (outgoing) proton, $q = p'-p$ is the momentum transfer, $\psi$ is the proton Dirac spinor, and $F^{(p)}_1$ and $F^{(p)}_2$ are the usual Dirac and Pauli form factors for the proton.
Let us consider the $\mu = 0$ matrix element.
The nonrelativistic expansion of the right hand side of Eq.~\eqref{eq:proton_matrix_element} is
    \begin{align}
        \bra{p'} \bar q Q_{\text{em}} \gamma^0 q \ket{p} & = 1 - \frac{1}{8 M_N^2} \vb q^2 \left[ F^{(p)}_1(0) + 2 F^{(p)}_2(0) + 8 M_N^2 F^{(p)'}_1(0) \right] + \cdots \, \label{eq:proton_matrix_element_exp},
    \end{align}
where $F_1'(0) = d F_1(q^2)/d q^2 \, \big|_{q^2 = 0}$, and the dots are omitted higher order contributions.
We can reorganize this expression through the Sachs form factor
    \begin{align}
        G^{(p)}_E(q^2) & = F^{(p)}_1(q^2) + \frac{q^2}{4 M_N^2} F^{(p)}_2(q^2) \, ,
    \end{align}
to find the matching relation \cite{manoharHeavyQuarkEffective1997}
    \begin{align}
        c^{(p)}_D & = 1 + 8 M_N^2 G^{(p)'}_E(0) \, ,
    \end{align}
where $G_E'(0) = d G^{(p)}_E(q^2)/d q^2 \, \big|_{q^2 = 0}$.
The conventional definition of the proton charge radius, $r_p^2 = 6 \, G'^{(p)}_E(0)$, therefore has a clearer meaning in terms of a matching coefficient \cite{pinedaChiralStructureLamb2005}. 
A similar expression holds for the neutron with  $G^{(p)}_E$ replaced by $G^{(n)}_E$.

For a \gls{LLalpha} result, we can match $c_D^{(s)} = (c_D^{(p)} + c_D^{(n)})/2$ at tree-level to the recent lattice calculations of the radii \cite{djukanovicElectromagneticFormFactors2024, djukanovicPrecisionCalculationElectromagnetic2024} since the \gls{eft} result should reproduce the pure \gls{qcd} result at this order.
Then, we can make use of the one-loop beta function, which has been calculated in heavy quark effective theory in Refs.~\cite{bauerRenormalizationGroupScaling1998, blokSpectatorEffectsHeavy2005, finkemeierRenormalizingHeavyQuark1997, balzereitHeavyQuarkEffective1996, balkQuarkEffectiveTheory1994}. 
Taking the \gls{qed} limit gives
    \begin{align}
        \mu_U \frac{d c^{(p)}_D}{d \mu_U} & = - \frac{32}{3} \frac{\alpha}{4 \pi} \, , \\
        \mu_U \frac{d c^{(n)}_D}{d \mu_U} & = 0 \, .
    \end{align}
These can be reorganized into the isoscalar and isovector components in Eq.~\eqref{eq:single_nucleon_L},
    \begin{align}
        \mu_U \frac{d c^{(s)}_D}{d \mu_U}  = \mu_U \frac{d c^{(v)}_D}{d \mu_U} = - \frac{32}{6} \frac{\alpha}{4 \pi} \, .
    \end{align}
In terms of the subtraction velocity $\nu$, the solution is
    \begin{align}
        c^{(s)}_D(\nu) & = c^{(s)}_D\left( \frac{m_\pi}{M_N} \right) - 2 \log \left( \frac{\alpha(M_N \nu^2)}{\alpha(m_\pi^2/M_N)} \right) \, .
    \end{align}
At tree-level with respect to \gls{qed} or $O(e)$, the Darwin coefficient is RG invariant, so we fix the value of $c_D(m_\pi/M_N)$ to the recent lattice calculation \cite{djukanovicElectromagneticFormFactors2024, djukanovicPrecisionCalculationElectromagnetic2024}.
We have
    \begin{align}
        c_D^{(s)}\left( \frac{m_\pi}{M_N} \right) & = \frac{1}{2} \left[1 + \frac{4 M_N^2}{3} \left( r^2_{p, \text{lat}} + r^2_{n, \text{lat}} \right)\right].
    \end{align}

\subsection{Two-nucleon sector}
The neutron-proton potential is expanded as
    \begin{align}
        V_{pn} & = \sum_{v=-1} \sum_{\vb p', \vb p} V^{(v)}_{abcd} (\vb p', \vb p) p^\dagger_{\vb p', a} p_{\vb p, b} n^\dagger_{- \vb p', c} n_{-\vb p, d} \, ,
    \end{align}
where $v$ tracks the order in the velocity expansion of each coefficient, and $a$, $b$, $c$, and $d$ are spin indices for the proton and neutron.
The \gls{lo}, \gls{nlo}, and \gls{n2lo} potential coefficients in the S-wave are given by
    \begin{align}
        V^{(-1)}_{abcd} & = C_{0}^{(\threeS)} P^{(1)}_{ab, cd} + C_{0}^{(\oneS)} P^{(0)}_{ab, cd} \, , \\
        V^{(0)}_{abcd} & = \frac{1}{2} \left( \vb p'^2 + \vb p^2 \right) \left[ C_{2}^{(\threeS)} P^{(1)}_{ab, cd} + C_{2}^{(\oneS)} P^{(0)}_{ab, cd} \right] \, , \\
        V^{(1)}_{abcd} & = \frac{1}{4} \left( \vb p'^2 + \vb p^2 \right)^2 \left[ C_{4}^{(\threeS)} P^{(1)}_{ab,cd} + C_{4}^{(\oneS)} P^{(0)}_{ab, cd} \right] \, ,
    \end{align}
where the projection operators are given by
    \begin{equation}
    \begin{split}
        P^{(1)}_{ab, cd} & = \frac{1}{4} \left( 3 \delta_{ab} \delta_{cd} + \sigma^i_{ab} \sigma^i_{cd} \right) \, , \\
        P^{(0)}_{ab, cd} & = \frac{1}{4} \left( \delta_{ab} \delta_{cd} - \sigma^i_{ab} \sigma^i_{cd} \right) \, .
    \end{split}
    \label{eq:spin_projections}
    \end{equation}
As pointed out in Ref.~\cite{richardsonRadiativeCorrectionsRenormalization2024}, our definition of $C_4$ is a linear combination of $C_4 + \tilde C_4$ that appears in most of the \gls{pionless} literature.
Specifically, Refs.~\cite{chenNucleonnucleonEffectiveField1999,phillipsImprovingConvergenceEffective2000,lenskyForwardDoublyvirtualCompton2021,beaneHadronsNucleiCrossing2001,rupakPrecisionCalculation$to$2000} treat the part of the $O(p^4)$ potential proportional to $\vb p^2 \vb p'^2$ as an enhanced contribution while the $\vb p^4 + \vb p'^4$ contribution is relegated to subleading order.
However, when renormalizing the potential in a dimensional regularization scheme, one should use a basis of on-shell operators, i.e. the full momentum structure $(\vb p^2 + \vb p'^2)^2$ should be kept as a single operator.
Any off-shell ambiguity in the operators can always be removed through field redefinitions in favor of on-shell operators that are either higher order in the \gls{pionless} expansion or contain more fields.
This point was first emphasized in the context of \NN scattering in Ref.~\cite{flemingNNLOCorrectionsNucleonNucleon2000}.

\begin{figure}[t]
    \subfloat[]{%
    \includegraphics[width=0.5\textwidth]{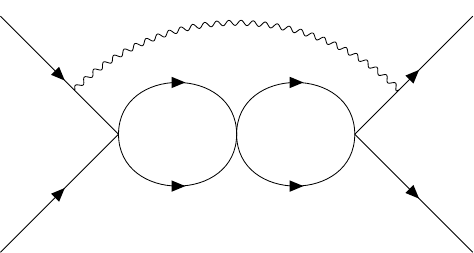}%
}
\caption{$O(\alpha/v)$ diagram that contributes to the anomalous dimension of the $C_2$ potential coefficient.}
\label{fig:c2_anomalous_dimension}
\end{figure}

\begin{figure}
    \centering
    \includegraphics[width = 0.75\textwidth]{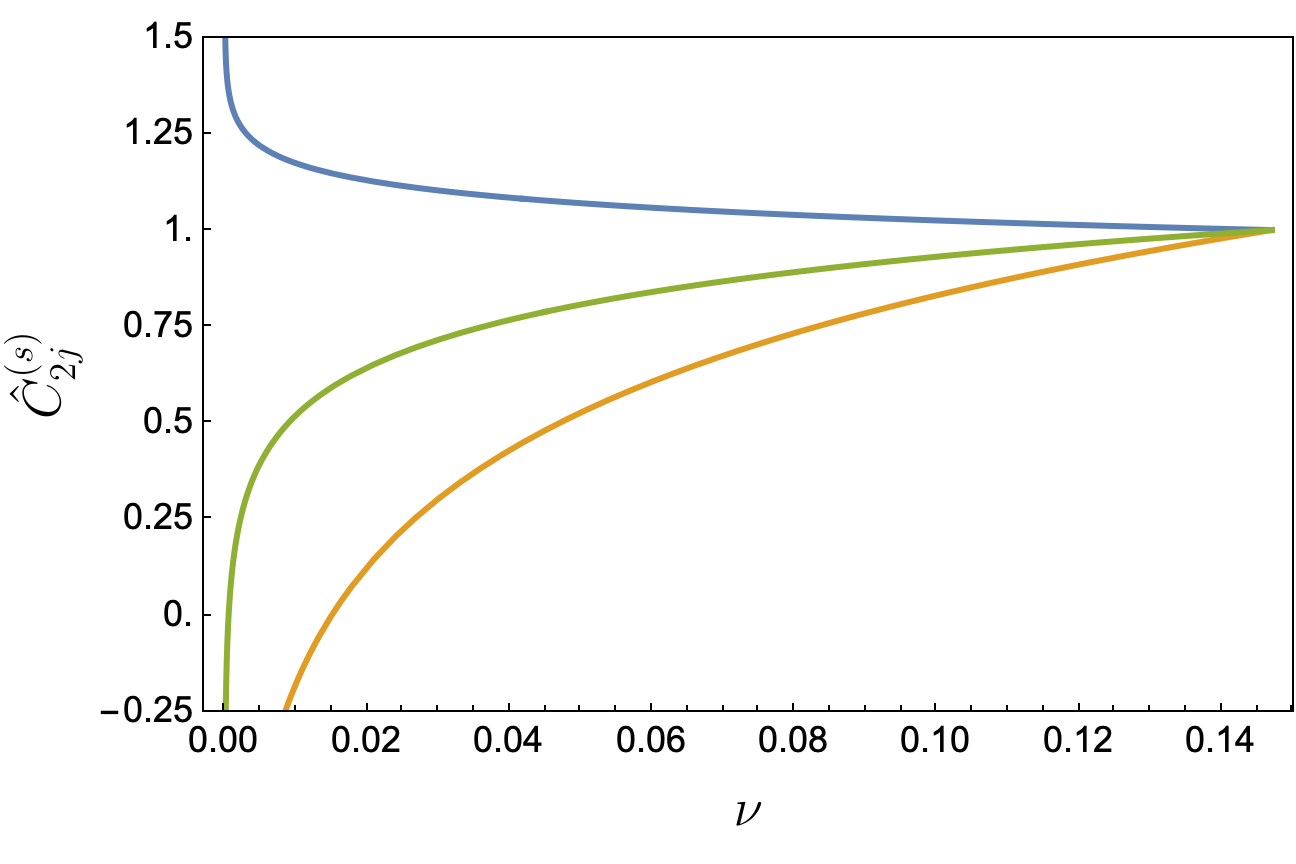}
    \caption{Running couplings normalized to their values at $\nu = m_\pi/M_N$. The blue line shows $\hat C_2^{(\threeS)}$, the orange line shows $\hat C_{4}^{(\threeS)}$, and the green line shows $\hat C_2^{(\oneS)}$.} \label{fig:running_lecs}
\end{figure}

In Ref.~\cite{richardsonRadiativeCorrectionsRenormalization2024}, we calculated the anomalous dimensions of $C_0$, $C_2$ and $C_4$ in the $\threeS$ channel.
The analysis in the $\oneS$ channel is identical.
The anomalous dimensions arise from diagrams such as the one shown in Fig.~\ref{fig:c2_anomalous_dimension}, which gives the anomalous dimension of $C_2$.
In the figure, there is an $A^0$ photon minimally coupled to the external proton lines with two intermediate two-nucleon bubbles.
In general, the $C_{2j}$ potential coefficient, where $j$ is a positive integer, is renormalized by a similar diagram with $2j$ bubbles.
In the $\oneS$ and $\threeS$ channels, the corresponding \gls{vRG} equations are\footnote{This corrects an error in the anomalous dimension of $C_0$ in Ref.~\cite{richardsonRadiativeCorrectionsRenormalization2024}; however, the main results of Ref.~\cite{richardsonRadiativeCorrectionsRenormalization2024} are unchanged since the running of $C_0$ was negligible.}
    \begin{align}
        \nu \frac{d C_0}{d \nu} & = 0 \, , \\
        \nu \frac{d C_{2j}}{d \nu} & = \frac{\alpha}{\pi} \frac{3 + 6j}{j+1} C_0 \left( \frac{ i M_N C_0 }{4 \pi} \right)^{2j} \, .
    \end{align}
The solutions are
    \begin{align}
        C_2(\nu) & = C_2\left( \frac{m_\pi}{M_N} \right) - \frac{27}{8} \left( \frac{M_N}{4 \pi} \right)^2 C_0^3 \log \left( \frac{\alpha(M_N \nu^2)}{\alpha(m_\pi^2/M_N)} \right) \, , \label{eq:running_c2} \\
        C_4(\nu) & = C_4\left( \frac{m_\pi}{M_N} \right) + \frac{15}{4} \left( \frac{M_N}{4 \pi} \right)^4 C_0^5 \log \left( \frac{\alpha(M_N \nu^2)}{\alpha(m_\pi^2/M_N)} \right) \label{eq:running_c4} \, .
    \end{align}
The values of the LECs at the matching scale $\nu = m_\pi/M_N$ where the pions are in principle integrated out are fixed according to,
    \begin{align}
        C^{(s)}_{0} & = \frac{4 \pi a_s}{M_N} \, , \label{eq:C0_matching} \\
        C^{(s)}_{2} \left( \frac{m_\pi}{M_N} \right) & = \frac{2 \pi a_s^2 r_s}{M_N} \, , \label{eq:C2_matching} \\
        C^{(s)}_{4} \left( \frac{m_\pi}{M_N} \right) & = \frac{4 \pi}{M_N} a_s^3 \left( \frac{1}{4} r_s^2 + \frac{P_s}{a_s} \right) \, , \label{eq:C4_matching}
    \end{align}
where $a_s$ and $r_s$ are the scattering length and effective range in channel $s$ taken from Ref.~\cite{wiringaAccurateNucleonnucleonPotential1995} with $\alpha = 0$, and $P_s$ is the shape parameter in channel $s$ taken from Ref.~\cite{deswartLowEnergyNeutronProtonScattering1995}.
The running LECs normalized to their values at $\nu = m_\pi/M_N$, i.e. $\hat C_{2j}(\nu) = \frac{C_{2j}(\nu)}{C_{2j}(m_\pi^2/M_N)}$, are shown in Fig.~\ref{fig:running_lecs}.

In the study of the deuteron electromagnetic properties, there are additional two-nucleon operators coupled to the electromagnetic field strength tensor.
There are the magnetic operators that typically appear at \gls{nlo} \cite{kaplanPerturbativeCalculationElectromagnetic1999, chenNucleonnucleonEffectiveField1999},
    \begin{align}
        \calL_{\NN B} & = e B^i \left[ L_1 \left( N^T P^i N \right)^\dagger \left( N^T \bar P^3 N \right) - i \epsilon^{ijk} L_2 \left( N^T P^j N \right)^\dagger \left( N^T P^k N \right) \right] + \text{H.c.} \, .
    \end{align}
Additionally, there is a one-derivative gauge invariant operator coupled to an electric field $\vb E$ \cite{lenskyForwardDoublyvirtualCompton2021},
    \begin{align}
        \calL_{\NN E} & =  e \left( \del \cdot \vb E \right) L_{1,E} \left( N^T P^i N \right)^\dagger \left( N^T P^i N \right) \, .
    \end{align}
In Ref.~\cite{lenskyForwardDoublyvirtualCompton2021}, this operator was counted as \gls{n3lo}.
Here this operator is treated as \gls{n2lo} because of the way we treat the $C_4$ interaction as discussed below Eq.~\eqref{eq:spin_projections}.

On one hand, $\vb E$ and $\vb B$ can be external virtual fields in the potential regime, which is relevant for the calculation of the deuteron form factors.
In this case, the external fields can change the soft momentum label of the nucleon fields, which we have omitted in these operators to condense the notation.
On the other hand, the electromagnetic fields can be treated as ultrasoft fields when the outgoing photon is real as is the case in $np \to d \gamma$.
In this case, the electromagnetic field does not change the momentum labels of the fields.
In either case, the four-nucleon-one-photon operators under consideration behave like $C_0$ vertices because they do not introduce factors of the \NN relative momentum.
Therefore, they are not renormalized at the order we consider here.

All together, the Lagrangian we will work with is
    \begin{align}
            \label{eq:nrqed_pionless:lagrangian}
        \calL = & \sum_{\vb p} N_{\vb p}^\dagger \left( i D_0 - \frac{\left( \vb p -i \vb D \right)^2}{2 M_N} \right) N_{\vb p} - \frac{1}{4} F_{\mu \nu} F^{\mu \nu} + \sum_{\vb p} \abs{p^\mu A_p^\nu - p^\nu A_p^\mu}^2 - \sum_{\vb p', \vb p} V(\vb p', \vb p)  \nonumber \\
        & - \frac{4 \pi \alpha}{2 M_N} \sum_{q, q', \vb p, \vb p'} \vb A_{q'} \cdot \vb A_{q} N^\dagger_{\vb p'} Q_N N_{\vb p} 
        + \frac{e}{2 M_N} \epsilon^{ijk} \left( \del^j A^k \right) \sum_{\vb p} N_{\vb p}^\dagger \sigma^i \left[ \kappa_0 + \kappa_1 \tau^3 \right] N_{\vb p} \nonumber \\
        & + \calL_{\NN E} + \calL_{\NN B}\, ,
    \end{align}
Power counting the velocity in diagrams is now straightforward. 
Nucleon and soft photon propagators count as $1/v^2$ while ultrasoft photon propagators count as $1/v^4$.
The purely \NN potentials follow the standard power counting of \gls{pionless}, where $Q \sim M_N v$ \cite{kaplanNewExpansionNucleonnucleon1998, kaplanTwoNucleonSystems1998, vankolckEffectiveFieldTheory1999}.
Finally, a soft loop has an integration measure that scales as $v^4$, a potential loop scales as $v^5$, and an ultrasoft loop scales as $v^8$.

Before proceeding to the electromagnetic properties of the deuteron, we comment on two related issues.
First, we review the estimate of where the diagrams with virtual photons like the one shown in Fig.~\ref{fig:c2_anomalous_dimension} enter in the \gls{pionless} power counting \cite{richardsonRadiativeCorrectionsRenormalization2024}.
This diagram as well as others with an arbitrary number of \NN bubbles inside the photon loop are $O(4 \pi \alpha/M_N p)$, where $p = M_N v \sim 40-50$ MeV in the bound state.
However, the expansion parameter of \gls{pionless} is usually taken to be $Q = p/m_\pi \sim 1/3$, which suggests that $\alpha \sim Q^4$ since $\alpha \sim 1/137$.
Therefore, we can conservatively estimate that the full contribution from diagrams such as the one shown in Fig.~\ref{fig:c2_anomalous_dimension} will enter at \gls{n3lo} at the earliest.
Second, these diagrams contain infrared divergences that would need to be canceled by real emission graphs.

Both of these issues may be circumnavigated by performing a \gls{vRG} improved calculation at \gls{n2lo}.
This amounts to using the running couplings in Eqs.~\eqref{eq:running_c2} and \eqref{eq:running_c4} with $\nu \sim 1/M_N a_s$, which resums the dominant \gls{LLalpha} effects from ultrasoft photons into $C^{(s)}_2$ and $C^{(s)}_4$.
Since we stop at \gls{n2lo}, we do not require the infrared divergent or non-logarithmic contributions from the diagrams.
The impact of neglecting the other contributions is instead estimated through the variation of the subtraction velocity $\nu$.
Additional details can be found in Appendix~\ref{app:logs}.


\section{Charge form factor}
    \label{sec:charge_ff}

The deuteron charge form factor has been studied extensively in \gls{pionless}.
It was calculated to \gls{nlo} in Ref.~\cite{kaplanPerturbativeCalculationElectromagnetic1999} and to \gls{n2lo} in Ref.~\cite{chenNucleonnucleonEffectiveField1999}.
Another analysis at \gls{n2lo} was performed with the development of the so-called $Z$-parameterization in Ref.~\cite{phillipsImprovingConvergenceEffective2000} and was recently extended to \gls{n3lo} in Ref.~\cite{lenskyForwardDoublyvirtualCompton2021}.
Here, we present a \gls{vRG} analysis of the charge form factor at \gls{n2lo}+\gls{LLalpha} without making use of the $Z$-parameterization.

The form factor is given by
    \begin{align}
        F_C(q^2) & = \frac{4 \gamma}{q} \tan^{-1} \left( \frac{q}{4 \gamma} \right) 
        - \frac{M_N \gamma^3}{2 \pi} C_2 \left[ 1 - \frac{4 \gamma}{q} \tan^{-1} \left( \frac{q}{4 \gamma} \right) \right] \nonumber \\
        & - \left( \frac{M_N \gamma^3}{2 \pi} C_2 \right)^2 \left[ 1 - \frac{4 \gamma}{q} \tan^{-1} \left( \frac{q}{4 \gamma} \right) \right] - \frac{M_N \gamma^5}{\pi} C_4 \left[ 1 - \frac{4 \gamma}{q} \tan^{-1} \left( \frac{q}{4 \gamma} \right) \right] \nonumber \\
        & - \frac{q^2 \gamma^3}{2 \pi} \left[\frac{1}{8} M_N C_4 - L_{1,E} \right] - \frac{1}{8 M_N^2} c_D q^2 \frac{4 \gamma}{q} \tan^{-1} \left( \frac{4 \gamma}{q} \right) \, . \label{eq:charge_ff_with_L}
    \end{align}
Recall that the \gls{lecs} apart from $L_{1,E}$ are functions of the subtraction velocity $\nu$.
The binding momentum $\gamma$ is related to the binding energy $B$ according to $\gamma = \sqrt{M_N B}$, and is given through \gls{n2lo} by \cite{richardsonRadiativeCorrectionsRenormalization2024},
    \begin{align}
        \gamma & = \frac{4 \pi}{M_N C_0} + \frac{M_N C_2}{4 \pi} \left( \frac{4 \pi}{M_N C_0} \right)^4 + \left[ \frac{3 M_N^2}{(4 \pi)^2} C_2^2 \left( \frac{4 \pi}{M_N C_0} \right)^7 - \frac{M_N}{4 \pi} C_4 \left( \frac{4 \pi}{M_N C_0} \right)^6  \right] \, .
    \end{align}
Even though $C_2$ and $C_4$ are running LECs, $\gamma$ is $\nu$ independent through \gls{n2lo} if the logarithms of $\alpha$ in Eqs.~\eqref{eq:running_c2} and \eqref{eq:running_c4} are expanded as,
    \begin{align}
        \log(\frac{\alpha(M_N \nu^2)}{\alpha(m_\pi^2/M_N)} ) & \approx - \frac{4 \alpha_{\text{OS}}}{3 \pi} \log \left( \frac{m_\pi \nu}{m_e} \right) \cdots \, .
    \end{align}
The residual dependence on $\nu$ in the resummed logarithms may be varied to probe the role of radiative corrections that are higher order in $\alpha$ (for a pedagogical discussion, see Ref.~\cite{Cohen:2019wxr}).

In Ref.~\cite{lenskyForwardDoublyvirtualCompton2021}, the contact term analogous to our $L_{1,E}$ was included at \gls{n3lo}.
As discussed in Sec.~\ref{sec:background}, this operator should be included at \gls{n2lo} due to the treatment of the four-derivative potential in an on-shell basis; this can be shown using the \gls{PDS} scheme \cite{kaplanNewExpansionNucleonnucleon1998}.
The PDS calculation in Ref.~\cite{lenskyForwardDoublyvirtualCompton2021} indicates that $L_{1, E}$ can be decomposed as
    \begin{align}
        L_{1,E} & = \frac{M_N}{8} C_4\left( \frac{m_\pi}{M_N} \right) \left[ 1 + \tilde L_{1,E} \right] \label{eq:L1E} \, .
    \end{align}
In the \gls{n3lo} calculation \cite{lenskyForwardDoublyvirtualCompton2021}, a dimensionless coefficient in rough correspondence with our $\tilde L_{1,E}$ was fit to the deuteron charge radius and found to be suppressed relative to the first term in Eq.~\eqref{eq:L1E}.
Therefore, we assume this contribution will be small here as well and should be considered higher order in the \gls{eft} expansion.
This amounts to dropping the term proportional to $\frac{1}{8} M_N C_4 - L_{1,E}$ in Eq.~\eqref{eq:charge_ff_with_L} so that the expression for the form factor reads,
    \begin{align}
        F_C(q^2) & = \frac{4 \gamma}{q} \tan^{-1} \left( \frac{q}{4 \gamma} \right) 
        - \frac{M_N \gamma^3}{2 \pi} C_2 \left[ 1 - \frac{4 \gamma}{q} \tan^{-1} \left( \frac{q}{4 \gamma} \right) \right] \nonumber \\
        & - \left( \frac{M_N \gamma^3}{2 \pi} C_2 \right)^2 \left[ 1 - \frac{4 \gamma}{q} \tan^{-1} \left( \frac{q}{4 \gamma} \right) \right] - \frac{M_N \gamma^5}{\pi} C_4 \left[ 1 - \frac{4 \gamma}{q} \tan^{-1} \left( \frac{q}{4 \gamma} \right) \right] \nonumber \\
        & - \frac{1}{8 M_N^2} c_D q^2 \frac{4 \gamma}{q} \tan^{-1} \left( \frac{4 \gamma}{q} \right) \, . \label{eq:charge_ff_no_L}
    \end{align}

\begin{figure}
    \centering    \includegraphics[width=0.75\textwidth]{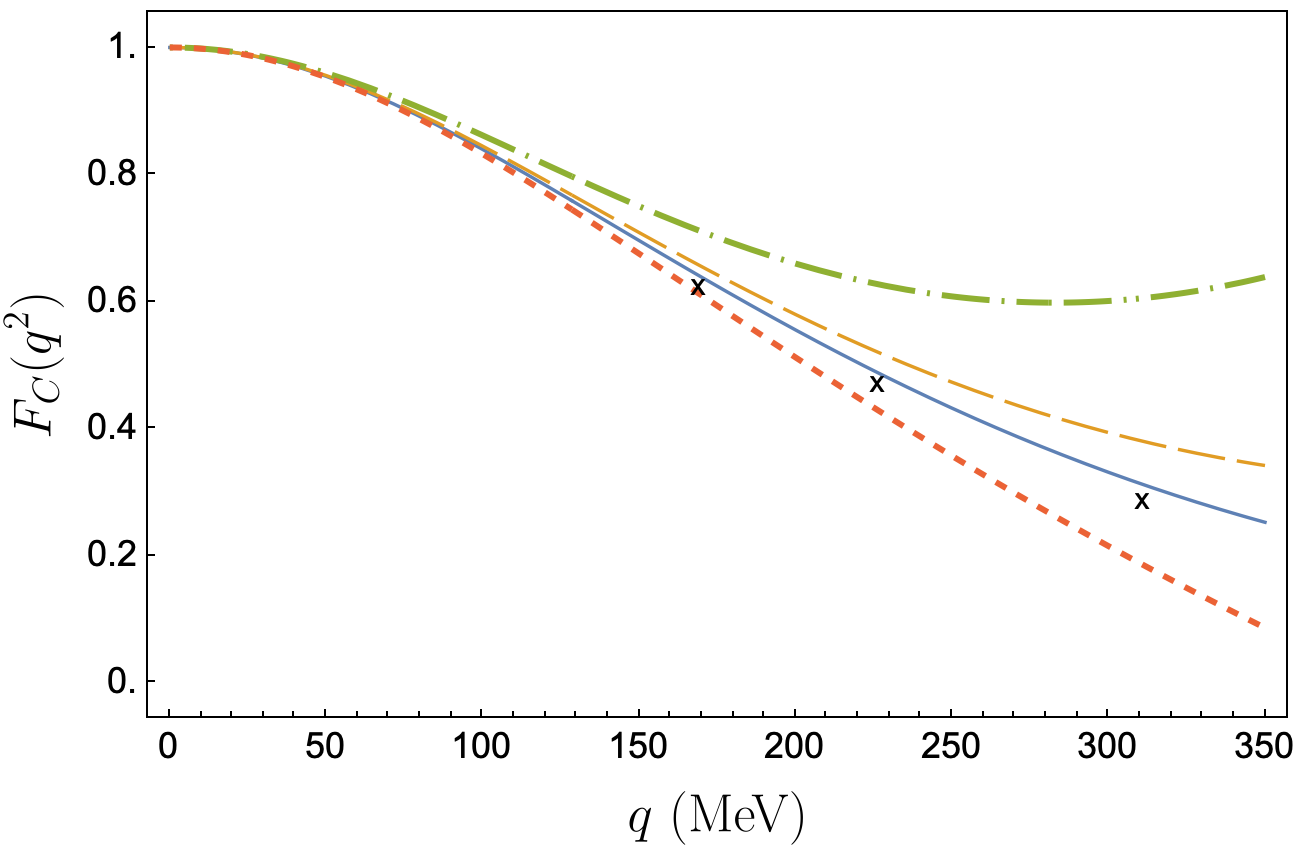}
    \caption{The solid (blue), dashed (orange), and dot-dashed (green) lines show the RG improved \gls{n2lo} charge form factor at $\nu = 0.1, \, 0.08,$ and $0.04$, respectively. The dotted (pink) line shows the \gls{n2lo} calculation in the $Z$-parameterization \cite{ lenskyForwardDoublyvirtualCompton2021}. The black crosses are data taken from Ref.~\cite{abbottPhenomenologyDeuteronElectromagnetic2000}.}
    \label{fig:charge_FF}
\end{figure}

The \gls{n2lo} RG improved result for the charge form factor is shown in Fig.~\ref{fig:charge_FF}.
For comparison, the pink dashed line shows the result of the $Z$-parameterization calculation at \gls{n2lo} \cite{ lenskyForwardDoublyvirtualCompton2021} and the experimental data is taken from Ref.~\cite{abbottPhenomenologyDeuteronElectromagnetic2000}.
We vary the subtraction velocity from $\nu = 0.1$ to $0.04$ as this provides an idea of the effect of higher order logarithms that would arise from higher order terms in $\alpha$ as well as the non-logarithmic contribution at $O(\alpha)$ in perturbation theory.
Interestingly, the $\nu = 0.1$ result is not far from the experimental data over a wide range of momentum transfers.
As $\nu$ is lowered even further, the theoretical prediction moves further away from the data, which might be improved in the next logarithmic order.
However, the momenta at which the deviation is most severe are well outside the radius of convergence of \gls{pionless}.
Additionally, we stress that the comparison to experimental data is not complete without calculating the full cross section with radiative corrections for, e.g., electron-deuteron scattering.
Here, we use this comparison as an indication that the \gls{vRG} is working properly.
On the other hand, this analysis \textit{is} complete at the order we consider insofar as the form factor is defined simply as the matrix element of the electromagnetic current operator.

\begin{figure}
    \centering
    \includegraphics[width=0.75\textwidth]{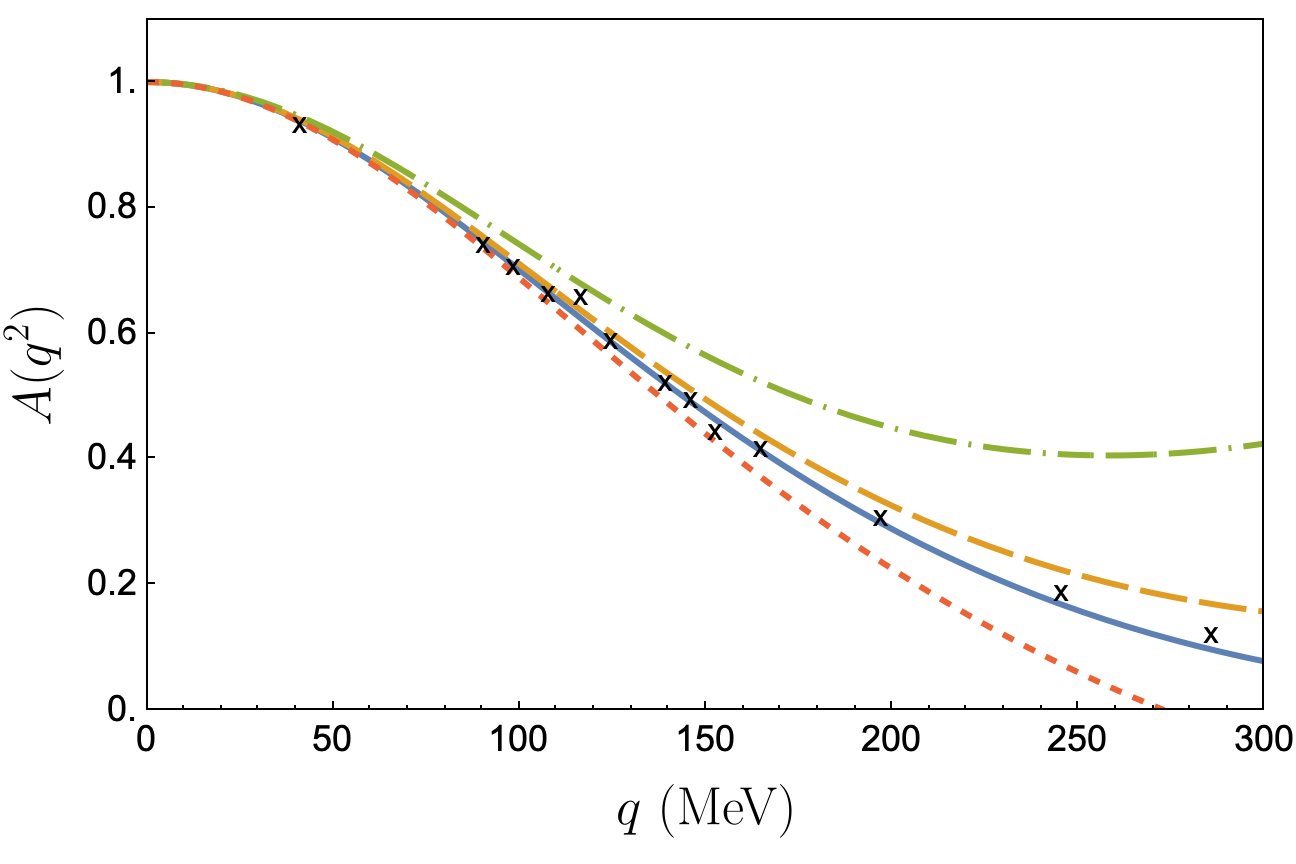}
    \caption{The solid (blue), dashed (orange), and dot-dashed (green) lines show the RG improved \gls{n2lo} charge form factor at $\nu = 0.1, \, 0.08,$ and $0.04$, respectively. The dotted (pink) line shows the \gls{n2lo} calculation in the $Z$-parameterization \cite{ lenskyForwardDoublyvirtualCompton2021}. The black crosses are data taken from Ref.~\cite{simonElasticElectricMagnetic1981}.}
    \label{fig:charge_A}
\end{figure}

In Fig.~\eqref{fig:charge_A}, we display the \gls{vRG} improved quantity,
    \begin{align}
        A(q^2) & = F_C^2(q^2) + \frac{2}{3} \eta F_M^2(q^2) + \frac{8}{9} \eta^2 F_Q^2(q^2) \, ,
    \end{align}
where $\eta = -q^2/4 M_d^2$, $F_M$ is the magnetic form factor, and $F_Q$ is the quadrupole form factor.
At the order we are working, we only need the magnetic form factor at \gls{lo} \cite{kaplanPerturbativeCalculationElectromagnetic1999},
    \begin{align}
        \frac{e}{2 M_d} F^{(0)}_M(q^2) & = \kappa_0 \frac{e}{M_N} F_C^{(0)}(q^2) \, .
    \end{align}
We compare the results to the analogous calculations in the $Z$-parameterization \cite{lenskyForwardDoublyvirtualCompton2021} at \gls{n2lo} and the data from Ref.~\cite{simonElasticElectricMagnetic1981}.
The features of the plot are identical to those in Fig.~\ref{fig:charge_FF}.
Again, we see excellent agreement with the data for $\nu = 0.1$ and $0.08$ for a wide range of momentum transfers.

From the form factor, we can also examine the conventional definition of the deuteron charge radius, 
    \begin{align}
        \langle r_d^2 \rangle_C & = - 6 \frac{d F_C(q^2)}{d q^2} \big|_{q^2 = 0} \, .
    \end{align}
At $\nu = m_\pi/M_N$, the theoretical charge radius $r_{\text{th}}(m_\pi/M_N) = \sqrt{\langle r_d^2 \rangle_C} = 2.15(6)$ fm, where we have assigned a $3\%$ error to account for the truncation of the \gls{eft}.
After the \gls{vRG} improvement, the charge radius at $\nu = 0.1$ is $r_{\text{th}}(0.1) = 2.11(6)$ fm.
Again, we have assigned a na\"ive 3\% error assuming that the \gls{vRG} improvement does not strongly modify the truncation error of the \gls{eft}.
Despite this uncertainty, the \gls{vRG} has shifted the central value by around 2\%.
A few representative values both with and without the truncation error estimate are shown in Table~\ref{tab:charge_radius}.

Because of the estimate for the truncation error, the theoretical result is still not as precise as the result of the CREMA collaboration \cite{pohlLaserSpectroscopyMuonic2016}, $r_{\text{CREMA}} = 2.12562(13)_{\text{exp}} (77)_{\text{th}}$, but it does agree within the \gls{pionless} uncertainty for $\nu \in [0.06, 0.1]$.
This should be compared to the theoretical value at the hard scale which is within a few-
percent of the experimental result.
Therefore, the summation of \gls{qed} logarithms through the \gls{vRG} seems to play an important role in reproducing the charge radius without recourse to the Z-parameterization, which subsumes corrections to all orders in $\alpha$ in the \gls{lecs}.

\begin{table}
    \centering
    \setlength{\tabcolsep}{4pt}
    \begin{tabular}{c  c c | c | c }
         & & $\nu = \frac{m_\pi}{M_N}$ & $\nu = 0.1$ & $\nu = 0.06$  \\
         \hline \hline
         $r_d$ (fm) &  no truncation error & 2.154(6) & 2.111(6) & 2.049(7) \\
         & with truncation error & 2.15(6) & 2.11(6) & 2.05(6) \\
         \hline \hline
    \end{tabular}
    \caption{Renormalization group improved predictions for the deuteron charge radius at $\nu = m_\pi/M_N$, 0.1, and 0.06.}
    \label{tab:charge_radius}
\end{table}

This result, however, should not be over-interpreted since this particular definition of $r_{\text{th}}$ depends explicitly on the subtraction velocity and is therefore not strictly an observable from the \gls{eft} point of view.
However, it is RG invariant to \gls{n2lo} if the logarithms $\alpha$ at different scales are re-expanded as discussed in the case of the binding momentum.
As mentioned at the end of Sec.~\ref{sec:background}, if we go to \gls{n3lo}, we should include the full contribution from \NN vertices dressed by an ultrasoft photon and not just the logarithmic contributions.
Then we will find infrared divergences in the form factor that would only cancel against real emission in, e.g., the differential cross section for electron-deuteron scattering.
However, if we were to match \gls{pionless} onto a new \gls{eft} consisting of point-like deuterons, nonrelativistic muons, and light electrons, then these IR divergences would necessarily cancel in the matching relations.
This step would also be relevant for providing a consistent description of radiative corrections in a uniform framework for very-low energy electron-deuteron scattering similar to previous analyses of lepton-proton scattering in the potential \gls{nrqed} framework \cite{pesetProtonRadiusPuzzle2021}.
We will explore this possibility in future work.

\section{Radiative neutron capture}
    \label{sec:npdgamma}

Lastly, we consider the radiative capture process $np \to d \gamma$, which is of fundamental importance to \gls{BBN}. 
This reaction is the first step in the BBN network and is the primary source of primordial deuterium.
A rigorous quantification of this reaction's theory errors is therefore imperative.
Simulations of \gls{BBN}~\cite{pisantiPArthENoPEPublicAlgorithm2008, pitrouPrecisionBigBang2018,burlesSharpeningPredictionsBigBang1999} use theory predictions from Refs.~\cite{rupakPrecisionCalculation$to$2000,andoRadiativeNeutronCapture2006} as input since the available experimental data are scarce and do not come with competitive errors. 
Reference~\cite{rupakPrecisionCalculation$to$2000} quotes a 1\% uncertainty in the energy regime relevant for BBN. 
In the following, we will investigate the sensitivity of this process to \gls{qed} driven radiative corrections.

In $np \to d \gamma$, the outgoing photon is real and therefore on-shell.
Thus, the photon must live in the ultrasoft regime, which greatly simplifies the calculation in the velocity framework.
Previous studies have calculated the cross section for this process without separating the photon into multiple modes.
In those works, the loop integrals were calculated with a transfer of energy and momentum from the nucleons.
Because of the multipole expansion for the ultrasoft modes in this \gls{eft}, there is only energy carried away and not three-momentum.
Therefore, we can calculate the cross section easily in the center-of-mass frame of the incoming nucleons.

\begin{figure}
    \centering
    \subfloat[\label{fig:npdgamma_loglog_noL}]{\includegraphics[width=0.49\textwidth]{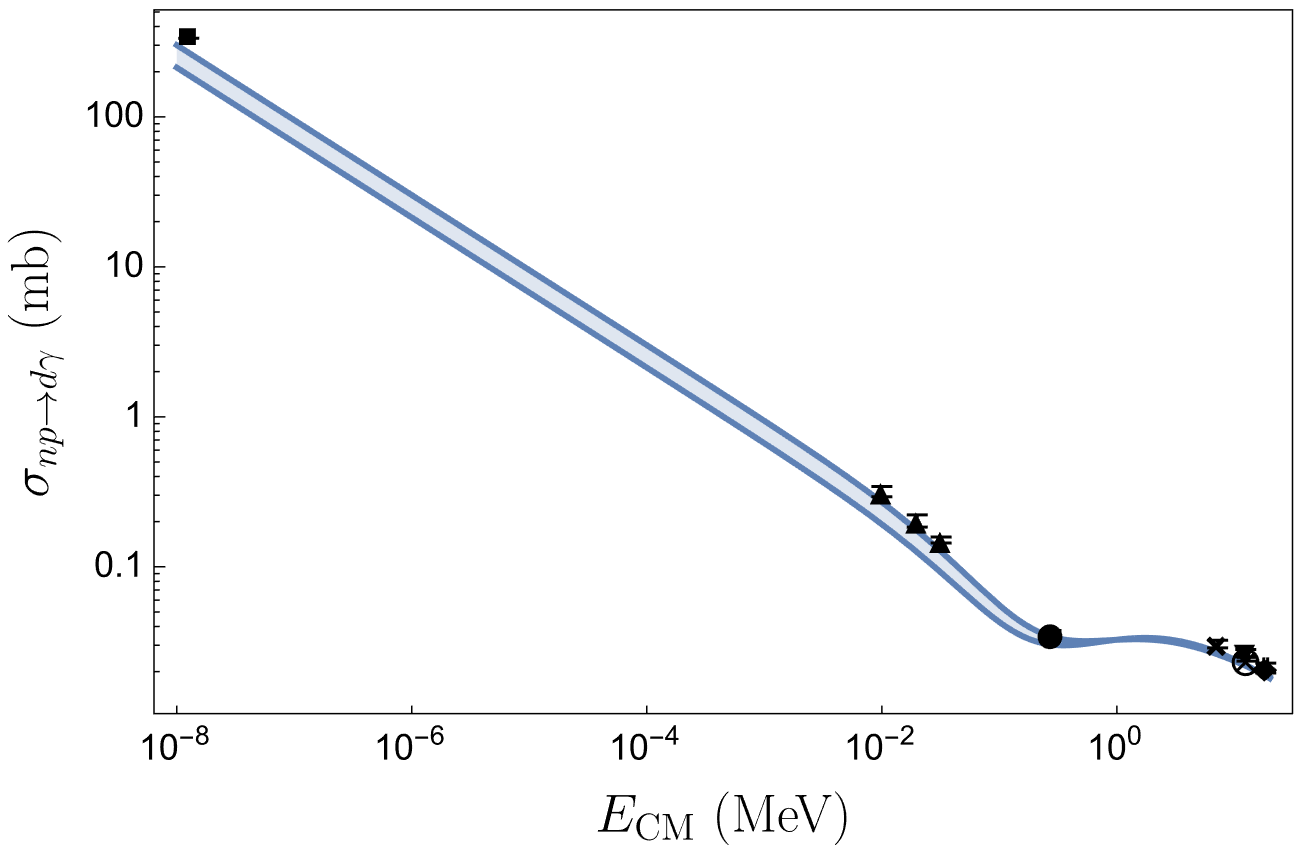}}
    \hfill
    \subfloat[\label{fig:npdgamma_loglog_withL}]{\includegraphics[width=0.49\textwidth]{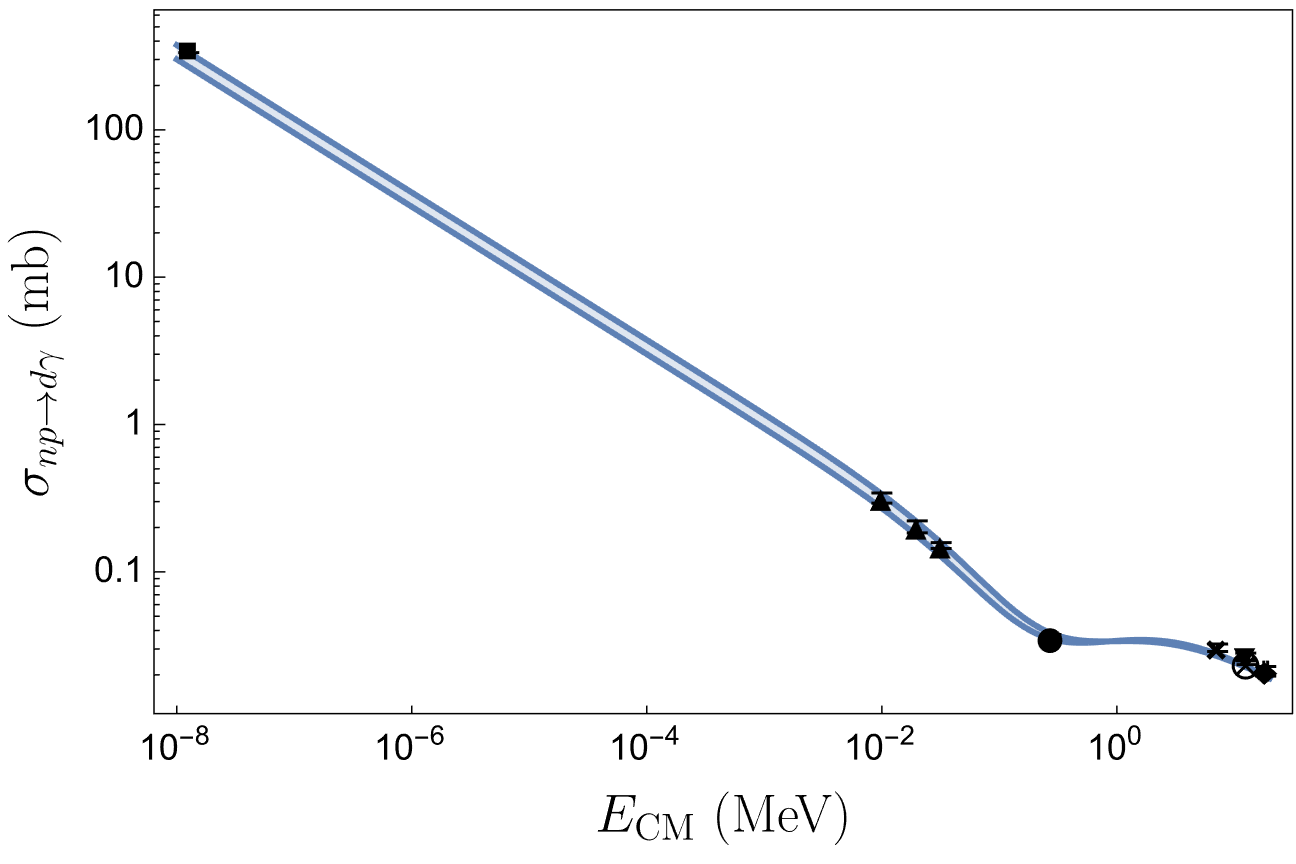}}
    \caption{Results for the radiative capture cross section (a) without and (b) with $\tilde L_1$. The band shows the variation of the subtraction velocity from $\nu = m_\pi/M_N$ to $0.01$. The experimental data come from Ref.~\cite{coxProtonthermalNeutronCapture1965} (square), \cite{suzukiFirstMeasurementPn1995} (upward triangles), \cite{nagaiMeasurement121997} (circle), \cite{tudoric-ghemoNeutronprotonRadiativeCapture1967} (cross), \cite{bosmanMeasurementTotalCross1979} (diamond), \cite{stiehlerTotalNpCapture1986} (downward triangle), and \cite{michelMeasurementCaptureReaction1989} (crossed circle).}  \label{fig:npdgamma_loglog}
\end{figure}

The amplitude is mostly driven by the M1 transition below center-of-mass-energy $E_{\text{CM}} = 0.1$ MeV and dominated by the E1 transition above this energy.
We reproduce the amplitude to \gls{nlo} \cite{chen$nstackrelensuremathrightarrowpdensuremathgamma$BigbangNucleosynthesis1999, rupakPrecisionCalculation$to$2000} using the renormalization group improved couplings.
The total cross section for this process to \gls{nlo} is given by
    \begin{align}
        \sigma_{np \to d \gamma} & = \frac{4 \pi \alpha_U \left( \gamma^2 + \vb p^2 \right)^3}{M_N^4 \gamma^3 \abs{\vb p} }  \left[ \abs{\tilde X_{M1}}^2 + \abs{\tilde X_{E1}}^2 \right] \, ,
    \end{align}
where $\vb p$ is the momentum of the neutron in the center-of-mass frame, $\gamma = \sqrt{M_N B}$ is the binding momentum of the deuteron, and $\tilde X_{M1}$ and $\tilde X_{E1}$ are the amplitudes defined in Refs.~\cite{chen$nstackrelensuremathrightarrowpdensuremathgamma$BigbangNucleosynthesis1999, rupakPrecisionCalculation$to$2000}.
We have also denoted the fine-structure constant here as $\alpha_U$ to make explicit that $\alpha_U = \alpha(M_N \nu^2)$ should be evaluated at the ultrasoft scale because the photon coupling is ultrasoft.
Through \gls{nlo}, the M1 amplitude is 
    \begin{align}
        \tilde X_{\text{M}1}^{\text{LO}} & = \kappa_1 \gamma^2 \left( \frac{1}{p^2 + \gamma^2} - \frac{M_N}{4 \pi} \frac{C_0^{(\oneS)}}{1 + i \frac{M_N p}{4 \pi} C_0^{(\oneS)} } \frac{1}{\gamma - i p}  \right) \, , \\
        \tilde X_{\text{M}1}^{\text{NLO}} & = \frac{M_N \gamma^3}{4 \pi} C_2^{(\threeS)} \left[\frac{\kappa_1}{2} \frac{1}{1 + i \frac{M_N p}{4 \pi} C_0^{(\oneS)}} + \tilde X_{\text{M}1}^{\text{LO}}  \right] \nonumber \\
        & + \frac{\kappa_1}{2} \frac{M_N \gamma^3}{4 \pi} C_2^{(\oneS)} \frac{1}{1 + i \frac{M_N p}{4 \pi} C_0^{(\oneS)}} \left( 1 - \frac{2 p^2}{\gamma (\gamma - i p)} \frac{1}{1 + i \frac{M_N p}{4 \pi} C_0^{(\oneS)}} \right) \nonumber \\
        & - \frac{M_N \gamma^3}{4 \pi} \frac{L_1}{1 + i \frac{M_N p}{4 \pi} C_0^{(\oneS)}}\, .
    \end{align}
The E1 amplitude is given by 
	\begin{align}
		\tilde X_{\text{E}1} & = - \frac{M_N p \gamma^2}{(p^2 + \gamma^2)^2} \left[1 + \frac{M_N \gamma^3}{4 \pi} C_2^{(\threeS)}  \right].
	\end{align}
Note that the multipole expansion of the ultrasoft interactions automatically generates the correct form of the amplitude \cite{chen$nstackrelensuremathrightarrowpdensuremathgamma$BigbangNucleosynthesis1999, rupakPrecisionCalculation$to$2000} without performing additional expansions.
This is an important feature of the construction of the \gls{eft}; we have set up the Lagrangian in a way that the power counting is satisfied in the amplitudes without the need to expand any results.

\begin{figure}
    \subfloat[\label{fig:suzuki}]{\includegraphics[width=0.49\textwidth]{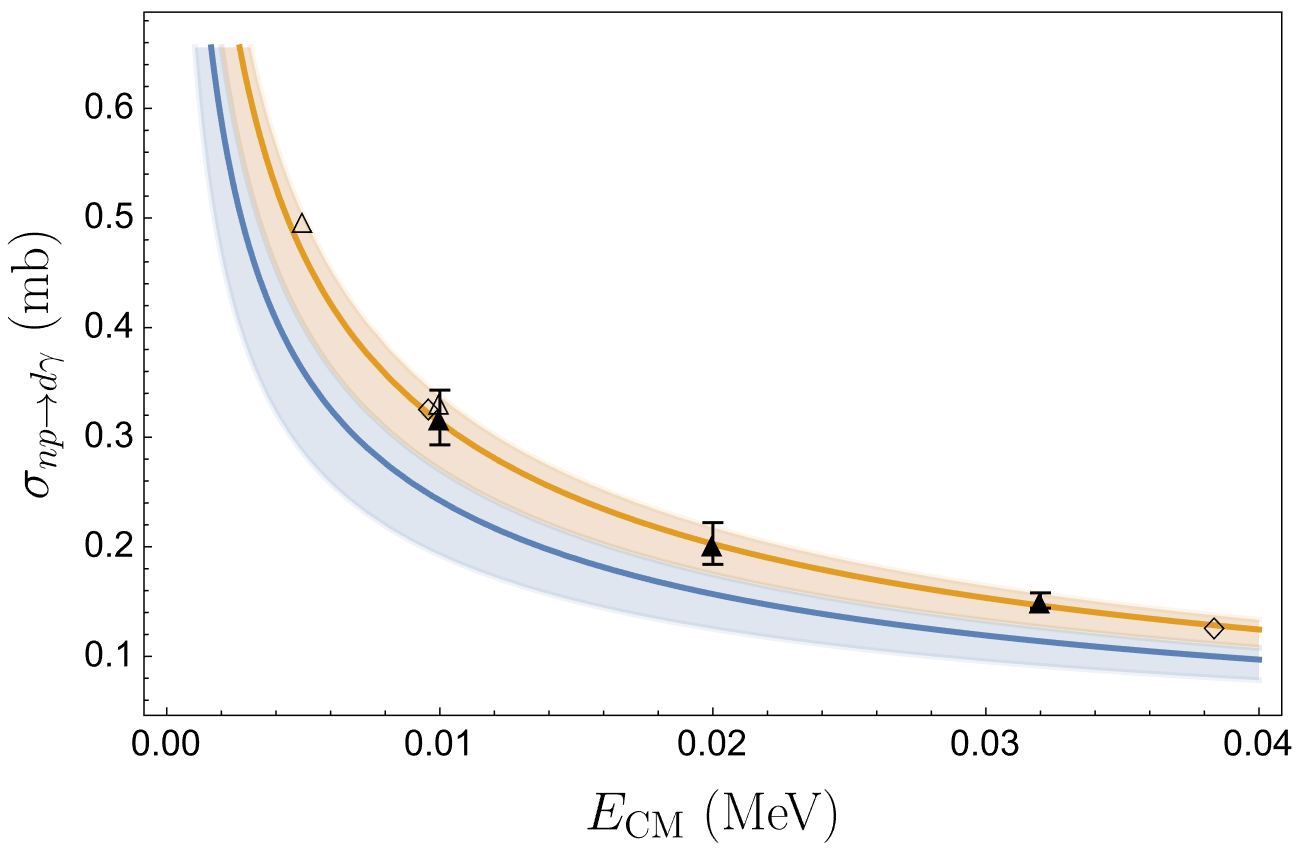}}
    \hfill
    \subfloat[\label{fig:nagai}]{\includegraphics[width=0.49\textwidth]{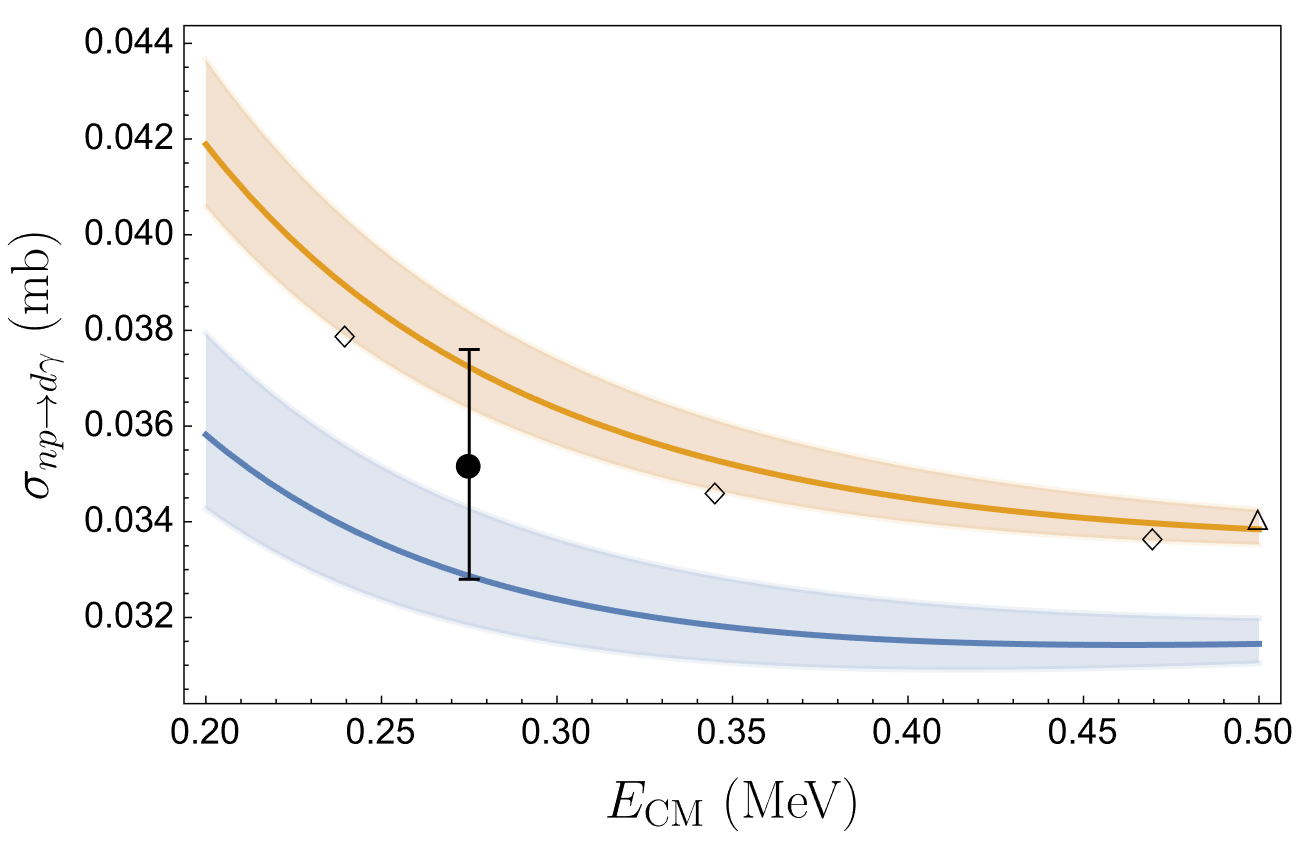}}
    \hfill
    \subfloat[\label{fig:slaus}]{\includegraphics[width=0.49\textwidth]{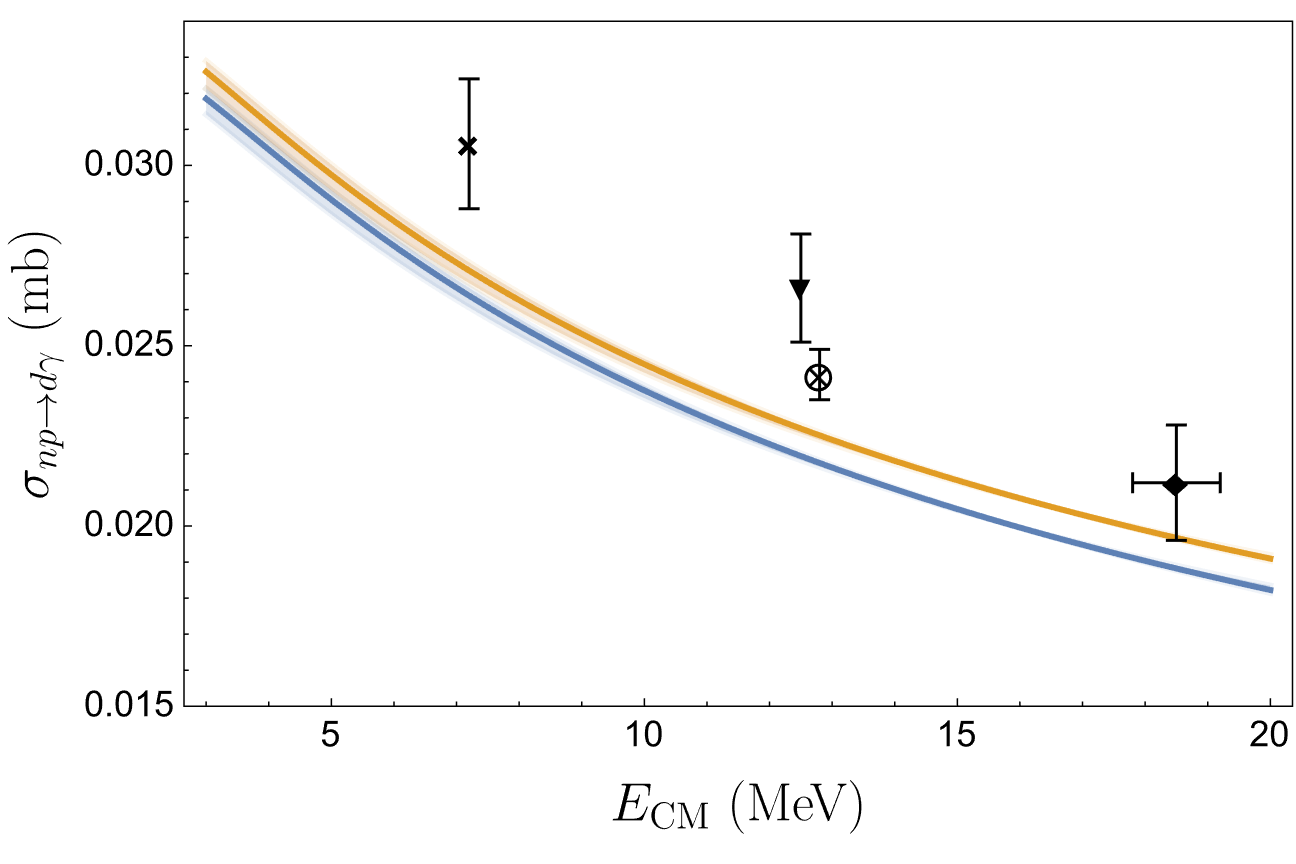}}
    \caption{Renormalization group improved cross section without $\tilde L_1$ in blue and with $\tilde L_1$ in orange. The open triangles are the \gls{pionless} result of Ref.~\cite{rupakPrecisionCalculation$to$2000}, and the open diamonds are the \gls{ChiEFT} result of Ref.~\cite{acharyaGaussianProcessError2022}. Other features are identical to Fig.~\ref{fig:npdgamma_loglog}.} \label{fig:npdgamma_bins}
\end{figure}

In a manner similar to $L_{1,E}$, we can use the PDS beta function \cite{chenNucleonnucleonEffectiveField1999} to tease out the parts of $L_1$ determined by \NN scattering parameters.
Then we take the PDS subtraction point to zero.
This results in 
    \begin{align}
            \label{eq:L1_PDS}
        L_1 & = \frac{\kappa_1}{2} \left( C_0^{(\oneS)} - C_0^{(\threeS)} \right) \left( \frac{C_2^{(\oneS)}\left( \frac{m_\pi}{M_N} \right)}{C_0^{(\oneS)}} - \frac{C_2^{(\threeS)}\left( \frac{m_\pi}{M_N} \right)}{C_0^{(\threeS)}} \right) + \tilde L_1 \, ,
    \end{align}
where $\tilde L_1$ is an integration constant. 
In Eq.~\eqref{eq:L1_PDS}, the $C_2$ couplings are evaluated at $\nu = m_\pi/M_N$ since this is the scale at which the \gls{eft} is defined. 
Moreover, $L_1$ is RG invariant at this order.
In order to have a true \gls{nlo}+\gls{LLalpha} result, $\tilde L_1$ should be matched to a pure \gls{qcd} result.
While this process was calculated on the lattice in Refs.~\cite{beaneInitioCalculationnpto2015, changMagneticStructureLight2015}, the differences in the underlying parameters make a direct comparison difficult.
Instead, we first consider the case that $\tilde L_1 = 0$, or at least suppressed relative to the first term in Eq.~\eqref{eq:L1_PDS}.
Then we follow the usual prescription and fit $\tilde L_1$ to the thermal capture cross section of Ref.~\cite{coxProtonthermalNeutronCapture1965} obtaining
    \begin{align}
            \label{eq:L1_extract}
        \tilde L_1 & = -613.4 \pm 0.5 \, \text{fm}^4 \, .
    \end{align}
The size of $\tilde L_1$ is due to the use of minimal subtraction and is in rough agreement with the PDS fit if the subtraction point in Ref.~\cite{chenNucleonnucleonEffectiveField1999} is taken to 0.
Relative to the first term in Eq.~\eqref{eq:L1_PDS}, $\tilde L_1$ is suppressed by about 1/5 and comes with the opposite sign.
The error in Eq.~\eqref{eq:L1_extract} comes purely from the experimental value of the cross section in Ref.~\cite{coxProtonthermalNeutronCapture1965}.

In Fig.~\ref{fig:npdgamma_loglog}, we show the total cross section with and without $\tilde L_1$.
The solid lines show the \gls{vRG} improved cross section with $\nu = 0.05$ and the bands show the variation from $\nu = m_\pi/M_N$ to 0.01.
If we go much below $\nu = 0.01$, the results start to become unstable, so we take this to be the lower bound of our analysis.
The data is taken from Refs.~\cite{coxProtonthermalNeutronCapture1965, suzukiFirstMeasurementPn1995, nagaiMeasurement121997, tudoric-ghemoNeutronprotonRadiativeCapture1967, bosmanMeasurementTotalCross1979, stiehlerTotalNpCapture1986, michelMeasurementCaptureReaction1989}.
In our approach, $\tilde L_1$ is still necessary to obtain agreement with experiment.
However, fitting $\tilde L_1$ alone is not enough to describe all of the data adequately.
This can be seen more clearly in Fig.~\ref{fig:npdgamma_bins} where we also compare our results to the \gls{pionless} \cite{rupakPrecisionCalculation$to$2000} and \gls{ChiEFT} results \cite{acharyaGaussianProcessError2022}.
In particular, the central values of the data from Ref.~\cite{suzukiFirstMeasurementPn1995} are almost reproduced with the presence of $\tilde L_1$ and $\nu = 0.05$.
In this region, running from $\nu = m_\pi/M_N$ to 0.05 induces a shift in the cross section of 6-8\%.
Both curves are within the error of the single data point from Ref.~\cite{nagaiMeasurement121997}, and the \gls{vRG} evolution leads to a change in the cross section of about 3\%.
At higher energies, the variation from the \gls{vRG} is negligible, so the higher order corrections in the \gls{pionless} power counting should be more important.

We find reasonable agreement with the \gls{pionless} \cite{rupakPrecisionCalculation$to$2000} and \gls{ChiEFT} results \cite{acharyaGaussianProcessError2022}.
However, we note that Ref.~\cite{rupakPrecisionCalculation$to$2000} states that electromagnetic effects, in particular the potential-like interaction coupling the proton and neutron magnetic moments, are included implicitly by making use of the experimental scattering length in the $\oneS$ channel.
A similar statement could be made regarding the \gls{ChiEFT} calculation since the parameters of the potential are generally fit to \NN scattering data.
In our approach, these effects are included via the renormalization group and go beyond the use of potential-like electromagnetic interactions.
Both Ref.~\cite{rupakPrecisionCalculation$to$2000} and Refs.~\cite{acharyaGaussianProcessError2022, acharyaUncertaintyQuantificationElectromagnetic2023} find sub-percent level precision, which is important for comparison with experiment and astrophysical observations and for validation of the general \gls{eft} approach.
However, a complete account of strong and electroweak effects will be required for this process to truly be considered a precision probe of the Standard Model and possible extensions. 
The renormalization group analysis presented here suggests that $np \to d \gamma$ is sensitive to \gls{qed} corrections to the \NN interaction at the few-percent level, which seems to be an unquantified source of uncertainty in existing calculations.


\section{Conclusion}
    \label{sec:conclusion}

The role of radiative corrections in nonrelativistic bound states is an important theoretical topic applicable for a wide variety of systems.
In order to use nonrelativistic light nuclei for high-precision experiments and cosmology tests of the Standard Model, it is important to understand these corrections.
Here, we have continued the work initiated in Ref.~\cite{richardsonRadiativeCorrectionsRenormalization2024} to address these effects in the two-nucleon system by performing a velocity renormalization group \cite{lukeRenormalizationGroupScaling2000} analysis of the deuteron charge form factor and the radiative neutron capture process $np \to d \gamma$.

For the charge form factor, we found that the \gls{vRG} evolution improves the \gls{pionless} description of the experimental data \cite{abbottPhenomenologyDeuteronElectromagnetic2000,simonElasticElectricMagnetic1981} at \gls{n2lo}+\gls{LLalpha} order without fitting higher order \gls{lecs} to experimental data or making use of the Z-parameterization \cite{phillipsImprovingConvergenceEffective2000}.
The only inputs are the effective range parameters of the \gls{av18} potential without the electromagnetic interaction \cite{wiringaAccurateNucleonnucleonPotential1995}, which we use as a proxy for a true \gls{qcd} result.
This also allowed us to provide a parameter free \gls{LLalpha} prediction for the deuteron charge radius, $r_{\text{th}}(\nu = 0.1) = 2.11(6)$ fm, which is a $3\%$ shift from the value at $\nu = m_\pi/M_N$.
This value is in agreement with the result of the CREMA collaboration \cite{pohlLaserSpectroscopyMuonic2016}, but is not as precise due to the truncation of \gls{pionless}.
However, this particular prediction should be refined by matching to another \gls{eft} where the deuteron is treated as a point-like particle, which will be explored in future work. 
This matching and subsequent running would also be appropriate for the proposed DRad experiment to extract of the deuteron charge radius from electron-deuteron scattering.

The radiative capture process $np \to d \gamma$ is the first step in the big bang nucleosynthesis chain \cite{wagonerSynthesisElementsVery1967, steigmanPrimordialNucleosynthesisPrecision2007, ioccoPrimordialNucleosynthesisPrecision2009, cyburtBigBangNucleosynthesis2016, aghanimPlanck2018Results2020, cookeOnePercentDetermination2018, burlesSharpeningPredictionsBigBang1999}.
Therefore, robust theoretical predictions play an important role in understanding astrophysical observations.
Our analysis is in general agreement with available experimental data as well as other calculations in \gls{pionless} \cite{chen$nstackrelensuremathrightarrowpdensuremathgamma$BigbangNucleosynthesis1999,rupakPrecisionCalculation$to$2000} and \gls{ChiEFT} \cite{acharyaGaussianProcessError2022}.
However, our approach is the first to our knowledge to incorporate the effects of radiative corrections through the renormalization group.
Although, we have still tacitly included electromagnetic effects by fitting the LEC $\tilde L_1$ to the measured thermal neutron capture cross section \cite{coxProtonthermalNeutronCapture1965}.
The scale variation in our \gls{vRG} analysis indicates that this process is sensitive to higher order \gls{qed} corrections at the level of a few percent.
Thus, \gls{qed} corrections appear to constitute an unquantified source of uncertainty in the accepted theoretical prediction of the cross section.
Furthermore, a reassessment of thermal neutron capture from lattice QCD would help disentangle the strong and electromagnetic effects of this process even more.

This work serves as a step forward to incorporating all relevant effects in low-energy, few-nucleon systems.
However, the analysis does suggest that the charge form factor, charge radius, and the $np \to d \gamma$ cross section receive percent level corrections from \gls{qed}, which have not been quantified until now.
A more rigorous analysis of the uncertainty due to the scale variation and the matching scale along the lines of Ref.~\cite{hoangTopantitopThresholdILC2014} is certainly warranted.
Also, extending this analysis to \gls{NLLalpha} would help increase the precision of the results.

We anticipate that other few-body systems could be subject to corrections at the few-percent level like the processes studied in this work.
In particular, proton-proton fusion will possibly receive larger corrections than indicated in a recent study \cite{combesRadiativeCorrectionsProtonproton2024}.
Furthermore, incorporating the renormalization group analysis in three- and four-nucleon systems, possibly with techniques developed in Ref.~\cite{phillipsNumericalRenormalizationUsing2000}, would be an interesting challenge.


\begin{acknowledgments}
    We would like to thank Sonia Bacca, Vadim Lensky, Emanuele Mereghetti and Ubirajara van Kolck for interesting discussions.
    Also, we thank Matthias Schindler and Roxanne Springer for critical feedback. ICR would like to thank the Mainz Physics Academy for financial support.
    This work was supported in part by the Deutsche Forschungsgemeinschaft (DFG) through the Cluster of Excellence ``Precision Physics, Fundamental Interactions, and Structure of Matter'' (PRISMA${}^+$ EXC 2118/1) funded by the DFG within the German Excellence Strategy (Project ID 39083149), by the NSF through cooperative agreement 2020275 (TRR), and by the DOE Topical Collaboration “Nuclear Theory for New Physics,” award No. DE-SC0023663 (TRR).
\end{acknowledgments}


\appendix

\section{Velocity EFT}
In a typical \gls{eft}, we count powers of momenta.
For relativistic \gls{eft}s this is suitable because the energy $E$ and the magnitude of the momentum $\abs{\vb P}$ are of the same size.
In \gls{pionless}, nucleons are nonrelativistic particles with momentum $\abs{\vb P} \ll M_N$ and will have energy $E \sim \vb P^2/M_N$ after removing $M_N$ through a heavy particle decomposition.
Instead of counting powers of $E$ or $\vb P$, it becomes more convenient to count powers of velocity $v = \vb P/M_N \ll 1$.
This leads to the two separate but correlated scales described in the text, namely, the soft scale $\vb P \sim M_N v$ and the ultrasoft scale $E \sim M_N v^2$.

As discussed in Sec.~\ref{sec:background}, when virtual photons are included in the theory without a mode separation, different powers of $v$ can mix in a single Feynman diagram, which is not desirable from the \gls{eft} point of view.
One way to avoid this is to use the full photon field and employ the method of regions \cite{benekeAsymptoticExpansionFeynman1998} on a diagram-by-diagram basis.
However, this can quickly become cumbersome.

\begin{figure}
    \centering
    \includegraphics[width=0.4\textwidth]{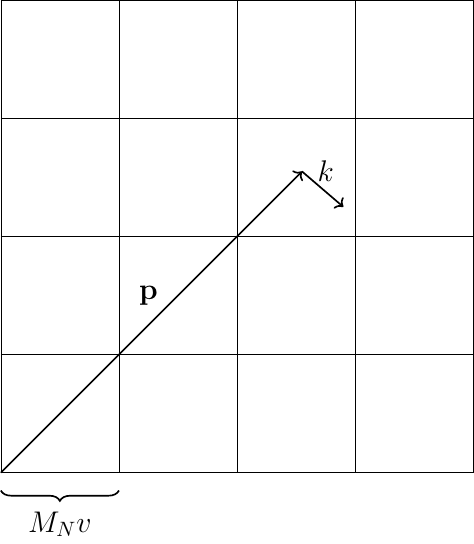}
    \caption{The binning of momentum space. Each bin has a typical size of $M_N v$. The soft momentum is labeled by $\vb p \sim O(M_N v)$ while the ultrasoft momentum is labeled by $k \sim O(M_N v^2)$.}
    \label{fig:momentum_bins}
\end{figure}

Instead, the velocity \gls{eft} framework allows us to reproduce the method of regions at the level of the Lagrangian.
The first step is to break up the nonrelativistic four-momentum $P$ of a nucleon as (cf. Eq.~\eqref{eq:background:nrqed:nucleon_momentum})
    \begin{equation*}
        (P^0, \vb P) = (0, \vb p) + (k_0, \vb k) \, ,
    \end{equation*}
where $\vb p$ is the momentum on the soft scale $M_N v$, $k_0$ is the energy on the ultrasoft scale $M_N v^2$, and $\vb k$ is a residual momentum on the ultrasoft scale.
This binning of momentum space, which is similar to the binning in heavy quark effective theory and heavy baryon \gls{ChiPT} can be visualized in Fig.~\ref{fig:momentum_bins}.
The nucleon field $N_{\vb p}$ carries a soft label $\vb p$ that indicates which bin the nucleon is in, and it carries residual ultrasoft momentum $k$ within the given $\vb p$ bin.
A soft photon field, $A_{p}^\mu$, behaves in the same way only with a soft four-momentum instead of a three-momentum.
An ultrasoft photon field, $A^\mu$, only carries ultrasoft momentum within the most bottom left block, which is called the zero bin \cite{manoharZeroBinModeFactorization2007}.

The momentum binning and separation of photon modes leads to helpful constraints on the interactions that can appear in the Lagrangian.
Due to energy conservation, there can not be a vertex containing an incoming and outgoing nucleon coupled to a single photon, i.e. $A_{\vb q} N^\dagger_{\vb p'} N_{\vb p}$.
If we allowed for a coupling of this type, then the outgoing nucleon would be far off-shell with (energy, momentum)$\sim (M_N v, M_N v)$ instead of $(M_N v^2, M_N v)$.
As a result, any vertex containing soft photons must have at least two photon legs. Therefore, the interaction between nucleons and soft photons starts at $O(\alpha)$.

\begin{table}
    \setlength{\tabcolsep}{18pt}
    \begin{tabular}{c|c|c}
            & (energy, momentum) & Degrees of freedom \\
            \hline
            potential & $(M_N v^2, M_N v)$ & $N_{\vb p}$   \\
            soft & $(M_N v, M_N v)$ & $A^\mu_{k}$ \\
            ultrasoft & $(M_N v^2, M_N v^2)$ &  $A^\mu$
    \end{tabular}
    \caption{Degrees of freedom and their velocity scalings.}
    \label{table:eft_dof}
\end{table}

\begin{figure}
    \centering
    \subfloat[\label{fig:ultrasoft_a}]{\includegraphics[width=0.49\textwidth]{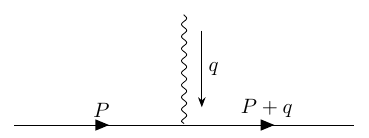}}
    \hfill
    \subfloat[\label{fig:ultrasoft_b}]{\includegraphics[width=0.49\textwidth]{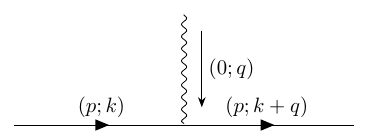}}
    \caption{(a) Photon-nucleon coupling without decomposing momenta into soft and ultrasoft components (b) Photon-nucleon coupling with the soft and ultrasoft decomposition.}  \label{fig:ultrasoft_interaction}
\end{figure}

On the other hand, a single ultrasoft photon is allowed to couple to the nucleon.
These interactions do not alter the soft label of the nucleon fields; the nucleon momentum can only shift within the same bin in Fig.~\ref{fig:momentum_bins} when interacting with an ultrasoft photon.
Thus, interactions with ultrasoft photons can only change the energy of nucleon propagators appearing in loops.
To see this, consider the kinetic term of the nucleon field,
    \begin{equation}
        \calL_{\text{kin}} =  \sum_{\vb p} N_{\vb p}^\dagger (x) \left( i \partial_0 - \frac{\vb p^2}{2 M_N} \right) N_{\vb p} (x) \, .
    \end{equation}
The derivative $\partial_0$ acts only on the residual component of the nucleon field while the $\vb p$ term can be though of as an operator that returns the soft label of the field.
This leads to the propagator,
    \begin{align}
        S(\vb p; k) = \frac{i}{k_0 - \frac{\vb p^2}{2 M_N} + i \epsilon} \, ,
    \end{align}
where we have explicitly separated the dependence on the soft three-momentum $\vb p$ and the ultrasoft four-momentum $k$ on the left-hand side.
The spatial part of $k$ never enters the nucleon propagator.
This is useful since it automatically implements the multipole expansion in Feynman diagrams containing virtual ultrasoft photons.
This can be illustrated by considering Fig.~\ref{fig:ultrasoft_interaction}, which could occur in a loop.
The first diagram shows the typical momenta we would expect without separating photon modes for an incoming nucleon with four-momentum $P$ and an incoming photon with momentum $q$.
In this case, the outgoing nucleon propagator would be given by
    \begin{align}
        S(P + q) & = \frac{i}{(P + q)^0 - \frac{(\vb P + \vb q)^2}{2 M_N} + i \epsilon} \, .
        \label{eq:nucleon_propagator}
    \end{align}
In a region where $P_0, q_0 \sim M_N v^2$, $\vb P \sim M_N v$, and $\vb q \sim M_N v^2$, the dependence on $\vb q$ can be expanded in a Taylor series.
Still, having to perform these expansions on a diagram-by-diagram basis is inconvenient.
The second diagram with the soft and ultrasoft momenta already decomposed leads to an outgoing nucleon propagator
    \begin{align}
        S(\vb p; k + q) & = \frac{i}{(k + q)^0 - \frac{\vb p^2}{2 M_N} + i \epsilon} \, ,
    \end{align}
which is just the first term in the expansion of Eq.~\eqref{eq:nucleon_propagator}.
Higher order terms in the expansion can be included through perturbative insertions of $N^\dagger_{\vb p} \, \vb p \cdot \del \, N_{\vb p}$ and $N^\dagger_{\vb p} \del^2 N_{
\vb p}$, where the gradient only acts on the residual components of $N$, as described below Eq.~\eqref{eq:single_nucleon_L}.

\section{Summing logarithms with the renormalization group}
    \label{app:logs}

In fixed-order perturbation theory in \gls{pionless}, a matrix element of an operator $\calO$, which we denote as $\langle \calO \rangle$, has an expansion of the form
    \begin{align}
        \langle \calO \rangle & = \calO_0 + \alpha \calO_1 + \alpha^2 \calO_2 + \cdots \, , 
        \label{eq:fixed_order}
    \end{align}
where $\calO_j$ is the sum of some subset of $O(\alpha^j)$ diagrams.
Each term also has an expansion in powers of the velocity $v$, but this detail is not important for this discussion.
Schematically, each term has the form,
    \begin{align}
        \calO_j = c_{j, 0} + c_{j, 1} \log \nu + c_{j, 2} \log^2 \nu + \cdots + c_{j, j} \log^j \nu \, .
        \label{eq:fixed_order_logs}
    \end{align}
where $c_{j, n}$ is a generic combination of couplings and other factors that appears at $O(\alpha^j)$ and the index $n$ indicates the power of the corresponding logarithm of the subtraction scale $\nu$.
The argument of the logarithm will also contain other factors, but these details are again not relevant at the moment.
Additionally, $c_{j, 0}$ will also include an infrared divergent piece.

Typically, the convergence of the expansion in Eq.~\eqref{eq:fixed_order} can be improved by resumming different orders of logarithms.
To see this, we first reorganize the matrix element as
    \begin{align}
        \langle \calO \rangle & = \sum_{j=0} c_{j, 0} \alpha^j + \sum_{j=1} c_{j, j} (\alpha \log \nu)^j + \sum_{j=2} c_{j, j-1}\alpha (\alpha \log \nu)^{j-1} + \cdots \, ,
        \label{eq:fixed_order_logs_reorg}
    \end{align}
where the first sum contains all non-logarithmic contributions, the second sum is the \gls{LLalpha} series, the third term is the \gls{NLLalpha} series and so on.
The resummation of a particular series is achieved by integrating the beta functions of the LECs.
For example, integrating the $O(\alpha)$ beta function of an LEC $C$ will resum the \gls{LLalpha} series into the running LEC $C(\nu)$.
Renormalization group improved perturbation theory then refers to using the running LEC in ordinary perturbation theory where the subraction velocity is taken to be on the order of the soft scale of the relevant system.
For example, when we use Eq.~\eqref{eq:running_c2} in perturbation theory, this sums up part of the second term of Eq.~\eqref{eq:fixed_order_logs_reorg} into $C_2$.
Expanding the resummed logarithms would reproduce the sum of $(\alpha \log \nu)^j$ terms from all $O(\alpha^j/v)$ diagrams with a single insertion of $C_2$.
A pedagogical introduction to this resummation can be found in Ref.~\cite{Cohen:2019wxr}.

\bibliography{radiative_references}
\end{document}